\documentclass[eng]{csam}
\usepackage{graphicx}
\usepackage{natbib}
\usepackage{amsmath,multirow}
\usepackage{mathtools}
\usepackage{enumerate,array}
\usepackage{float}
\usepackage{subfig} 
\usepackage{placeins}
\usepackage{placeins}

\submit{final}
\volumn{XX}{X}{XXXX}
\receive{0000}{00}{00} \revise{0000}{00}{00}  \accept{0000}{00}{00}
\setpagenum{1}

\heading{A General Class of New Continuous Mixture Distribution and Application}{ Brijesh P. Singh, Sandeep Singh \& Utpal Dhar Das}

\begin{document}

\title{A General Class of New Continuous Mixture Distribution and Application}

\author{ Brijesh P. Singh \address[a]{Department of Statistics, Institute of Science, Banaras Hindu University, Varanasi-221005, INDIA}, Sandeep Singh\footnote{Corresponding author: Research Fellow, Department of Statistics, Institute of Science, Banaras Hindu University, Varanasi-221005, INDIA. E-mail: sandeepsingh.stats@gmail.com}\same[a], Utpal Dhar Das\same[a]}

\begin{abstract}
A generalization of a distribution increases the flexibility particularly in studying of a phenomenon and its properties. Many generalizations of continuous univariate distributions are available in literature. In this study, an investigation is conducted on a distribution and its generalization. Several available generalizations of the distribution are reviewed and recent trends in the construction of generalized classes with a generalized mixing parameter are discussed. To check the suitability and comparobility, real data set have been used. 
\end{abstract}


\section{Introduction}
Modelling and analysis helps in explaining the lifetime events in various aspects of applied sciences. These phenomenon can be studied using various popular statistical distributions such as exponential, beta, gamma, pareto, weibull, lognormal etc. but each of these lifetime distributions has own advantages and disadvantages over one another due to the number of parameters involved, its shape, nature of hazard function and mean residual life function. Lindley (1958 and 1965) introduced one parameter continuous distribution as an alternative and competent distribution to the existing distributions with the help of convex combination of the particular case of gamma distribution. Lindley distribution is useful for analysing lifetime data, especially in stress-strength reliability modeling. 

\noindent Ghitany et al. (2008b) studied the properties of the one parameter Lindley distribution under a careful mathematical treatment. Shanker et al. (2015) made a comparison study of the goodness of fit of exponential and Lindley distributions on modeling of lifetime data. A generalized Lindley distribution, which includes as special cases the Exponential and Gamma distributions, was proposed by Zakerzadeh and Dolati (2009), and Nadarajah et al. (2011) introduced the exponentiated Lindley distribution. Ghitany and Al-Mutari (2008) considered a size-biased Poisson-Lindley distribution and Sankaran (1970) proposed the Poisson-Lindley distribution to model count data. Some properties of Poisson-Lindley distribution and its derived distributions were considered in Borah and Begum (2002) while Borah and Deka (2001a), Singh et al. (2015) considered the Poisson-Lindley and some of its mixture distributions. The zero-truncated Poisson-Lindley distribution and the generalized Poisson-Lindley distribution were considered in Ghitany et al. (2008a) and Mahmoudi and Zakerzadeh (2010), respectively.
 
\noindent A study on the inflated Poisson-Lindley distribution was presented in Borah and Deka (2001b), Singh et al. (2016) and Zamani and Ismail (2010) considering the Negative Binomial Lindley distribution. The weighted and extended Lindley distribution was considered by Ghitany et al. (2011), Elbatal and Elgarhy (2013) and Bakouch et al. (2012), respectively. The one parameter Lindley distribution in the competing risks scenario was considered in Mazucheli and Achcar (2011). The exponential Poisson Lindley distribution was presented in Barreto-Souza and Bakouch (2013). Ghitany et al. (2013) introduced the power Lindley distribution. Ali (2015) investigated various properties of the weighted Lindley distribution which main focus was the Bayesian analysis. Gomez-Deniz et al. (2014) studied Log Lindley distribution, another extended form of generalized Lindley distribution with applications to lifetime data is proposed by Torabi et al. (2014), Ashour and Eltehiwy (2015) considered exponentiated power Lindley distribution. A new four-parameter class of generalized Lindley distribution called the beta-generalized Lindley distribution is proposed by Oluyede and Yang (2015). 

\noindent Nedjar and Zeghdoudi (2016) suggested Gamma Lindley distribution and a generalized weighted Lindley distribution is discussed by Ramos and Louzada (2016). Kumar and Jose (2019) introduced a two-tailed version of the Lindley distribution through the name double Lindley distribution (DLD). Abouammoh et al (2015) proposed a generalized Lindley distribution and Tomy (2018) discussed several extension of Lindley distribution published at a glance. Ibrahim et al (2019) proposed a Topp Leone mixture of Generated Lindley (TLGLi), which is constructed based on the Topp Leone Generated (TLG) family introduced by Rezaei et al. (2017). Shanker and Amanuel studied quasi Lindley distribution and its properties. Shanker and Mishra (2013, 2014) introduced two parameter Lindley and its mixture with poisson distribution. Also three parameter Lindley distribution is considered by Shanker et al. (2017). Alkarni (2015) and  Pararai et al. (2015) discussed about power Lindley distribution and its extension. Mahmoudi and Zakerzadeh (2010) proposed an extended version of the compound Poisson-generalized Lindley distribution. Deniz and Ojeda (2011) proposed discretized version of Lindley distribution.

\noindent For exploration of flexibility of existing statistical distribution, generalization is also an interesting tool as well as important research area. Shanker and Shukla (2019) has proposed a two parameter life time distribution named Rama-Kamlesh distribution which is a generalised form of Lindley distribution containing various one parameter life time distributions (such as Lindley, Akash, Prakaamy distribution etc.). Again, Shukla et al. (2019) has proposed another generalisation named Shukla distribution which contains many other one parameter life time distribution like exponential, Shanker, Ishita, Pranav, Rani and Ram-Awadh distributions as particular cases of Shukla distribution. The aim of this study is to propose another generalized distribution which provides a more flexible distribution for modeling lifetime data. We propose a generalised distribution mixing gamma (2, $\theta$)  with gamma ($n$+2, $\theta$)  with different mixing parameters.

\noindent Lindley distribution is a mixture of gamma (1, $\theta$) i.e. exponential distribution ($\theta$) with gamma (2, $\theta$) as given
\begin{equation}
	f(x;\theta)=pf_{1}(x)+(1-p)f_{2}(x)\nonumber
\end{equation}
where
\begin{align}
	 p=\frac{\theta}{\theta+1} ;\quad f_{1}(x)=\theta e^{-\theta x}, \quad
	 f_{2}(x)=\frac{\theta^2}{\Gamma(2)}xe^{-\theta x}\nonumber
\end{align}
\noindent Thus the probability density function (pdf) of one parameter Lindley distribution is given by  \newline
\begin{align}
	f_1(x,\theta)=\frac{\theta^2}{1+\theta}(1+x) e^{-\theta x} \quad x>0,\quad\theta>0\nonumber
\end{align}
\noindent Correspondingly to above pdf, the cumulative distribution function (cdf) is given as
\begin{align}
	F_1(x,\theta)=1-\left(1+\frac{\theta x}{1+\theta}\right)e^{-\theta x}
\end{align}
\noindent For a random variable with the one parameter Lindley distribution, the probability distribution function is unimodal for $0<\theta<1$ and decreasing when $\theta>1$. The hazard rate function is an increasing function in $t$ and $\theta$ and given by
\begin{equation}
	h_1(t)=\frac{\theta^2(1+t)}{\theta+1+\theta t};\quad \theta>0,t>0\nonumber
\end{equation}

\noindent The generalised Lindley with two parameters named Rama-Kamlesh distribution (RKD) is mixture of exponential ($\theta$) and gamma ($\alpha$, $\theta$) and Shukla distribution (SD) is mixture of exponential ($\theta$) with gamma ($n$+1, $\theta$).
 
\begin{figure}[H]
	\centering
	\subfloat{{\includegraphics[width=6.8cm]{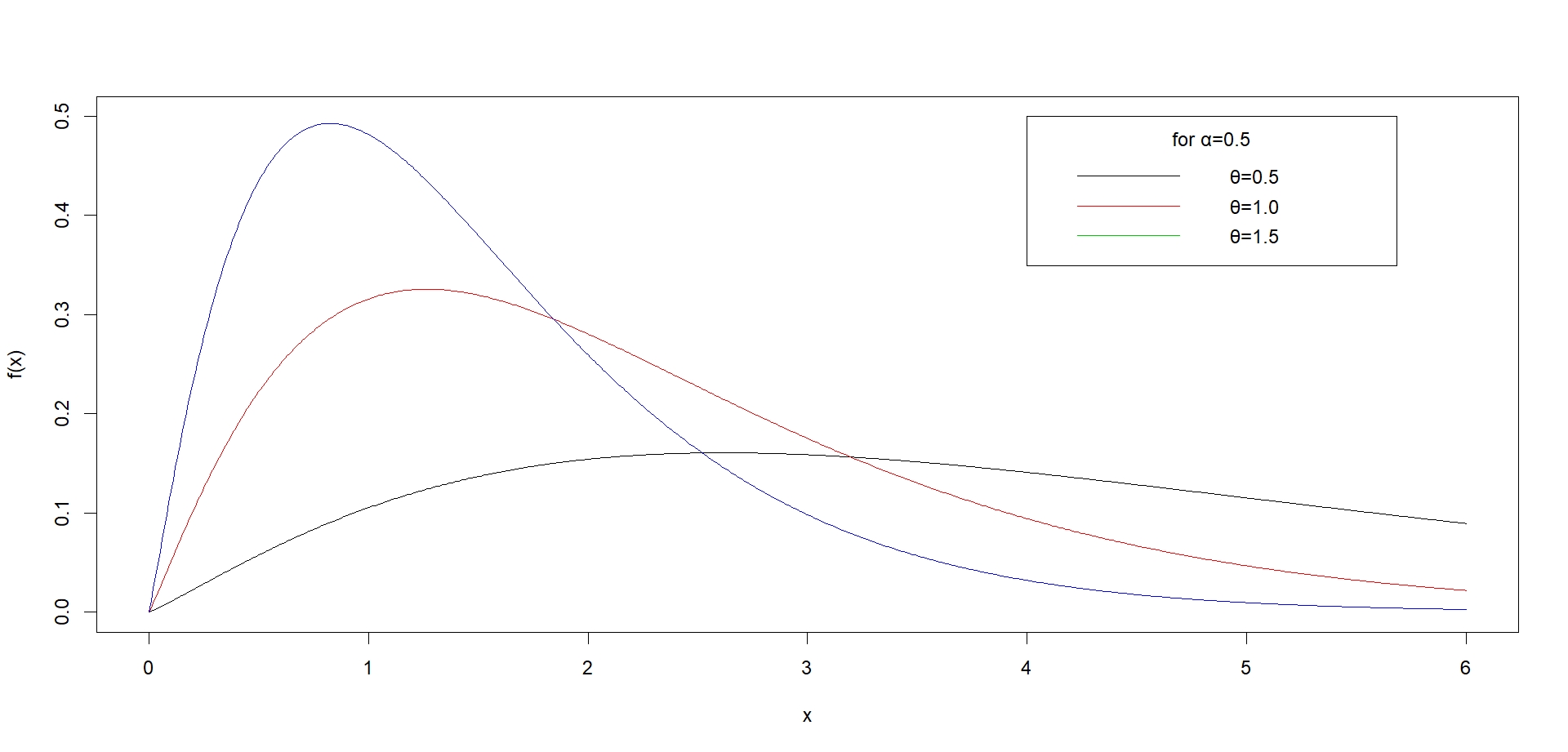}}}%
	\quad
	\subfloat{{\includegraphics[width=6.8cm]{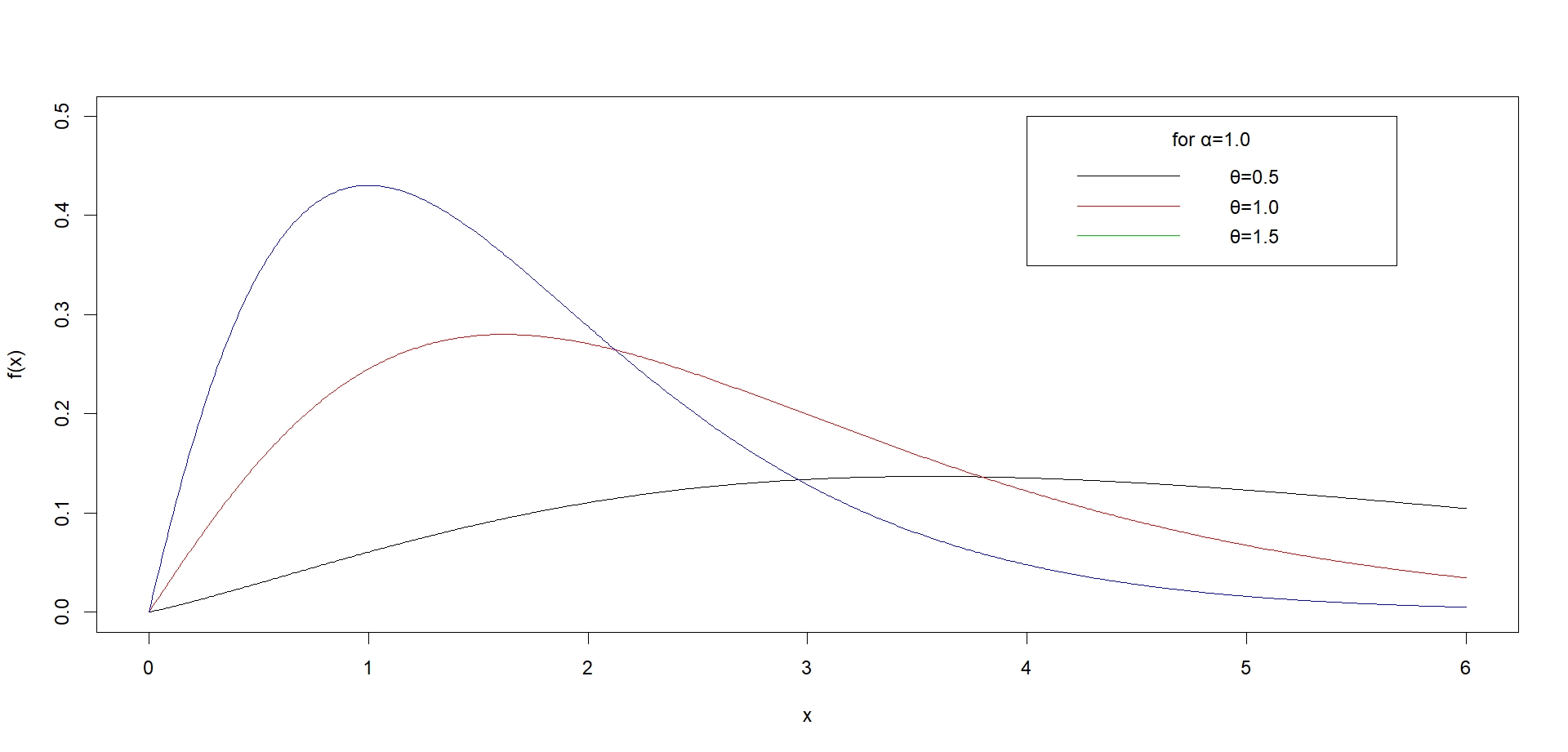}}}%
	\quad
	\subfloat{{\includegraphics[width=6.8cm]{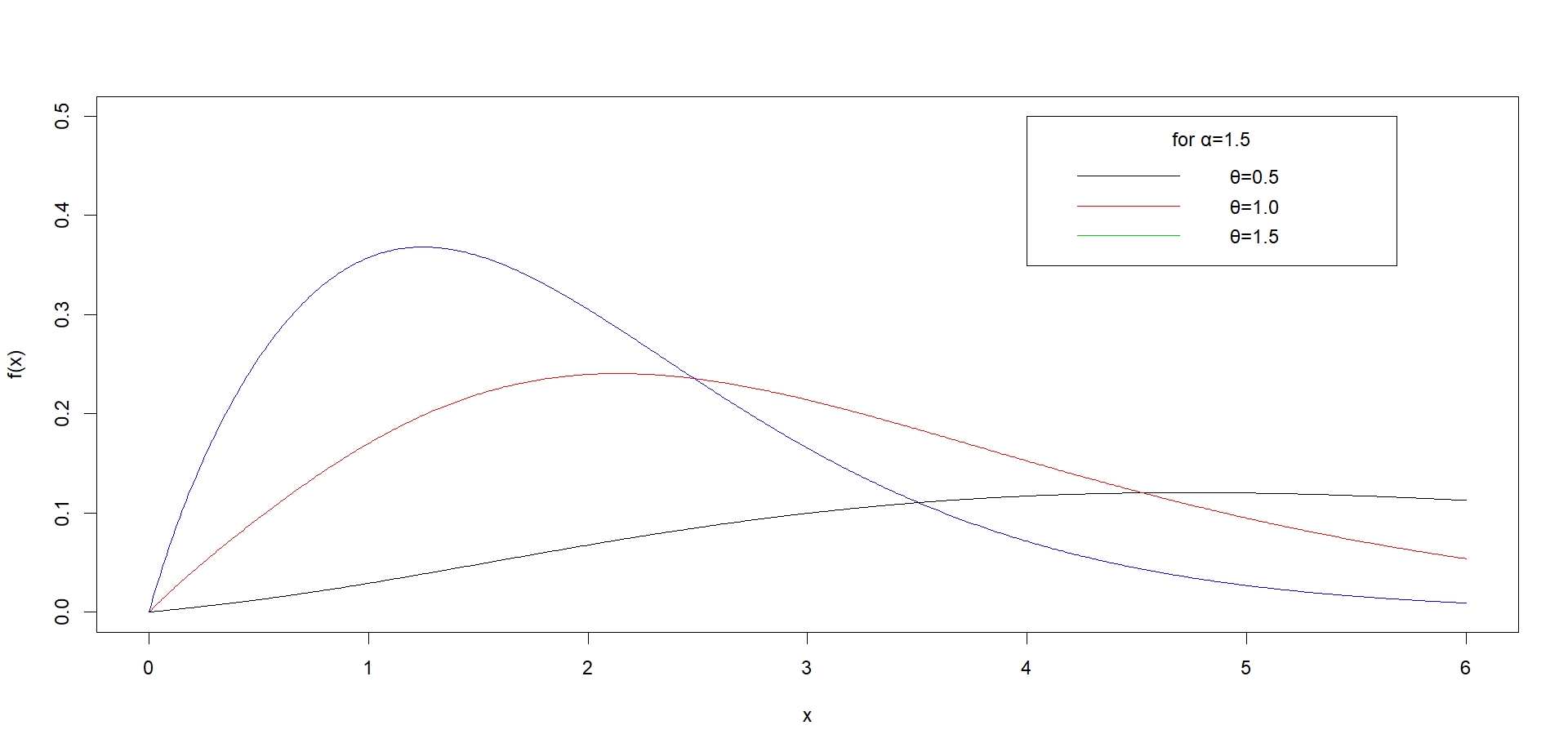}}}%
	\quad
	\subfloat{{\includegraphics[width=6.8cm]{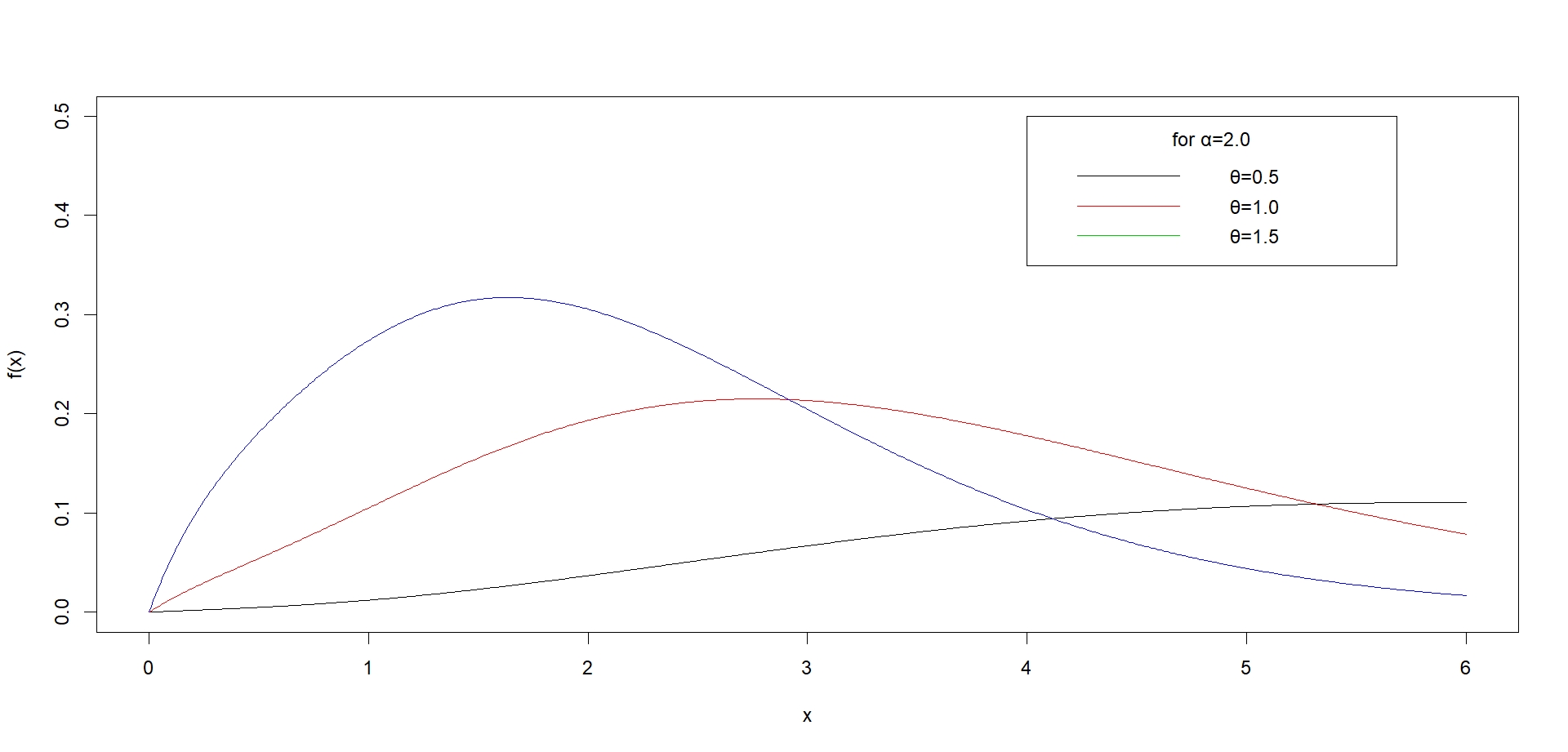}}}%
	\caption{pdf plot of SSD distribution for different values of $\alpha$ and $\theta$}
	\label{fig:pdf}
\end{figure}

\begin{figure}[H]
	\centering
	\subfloat{{\includegraphics[width=6.8cm]{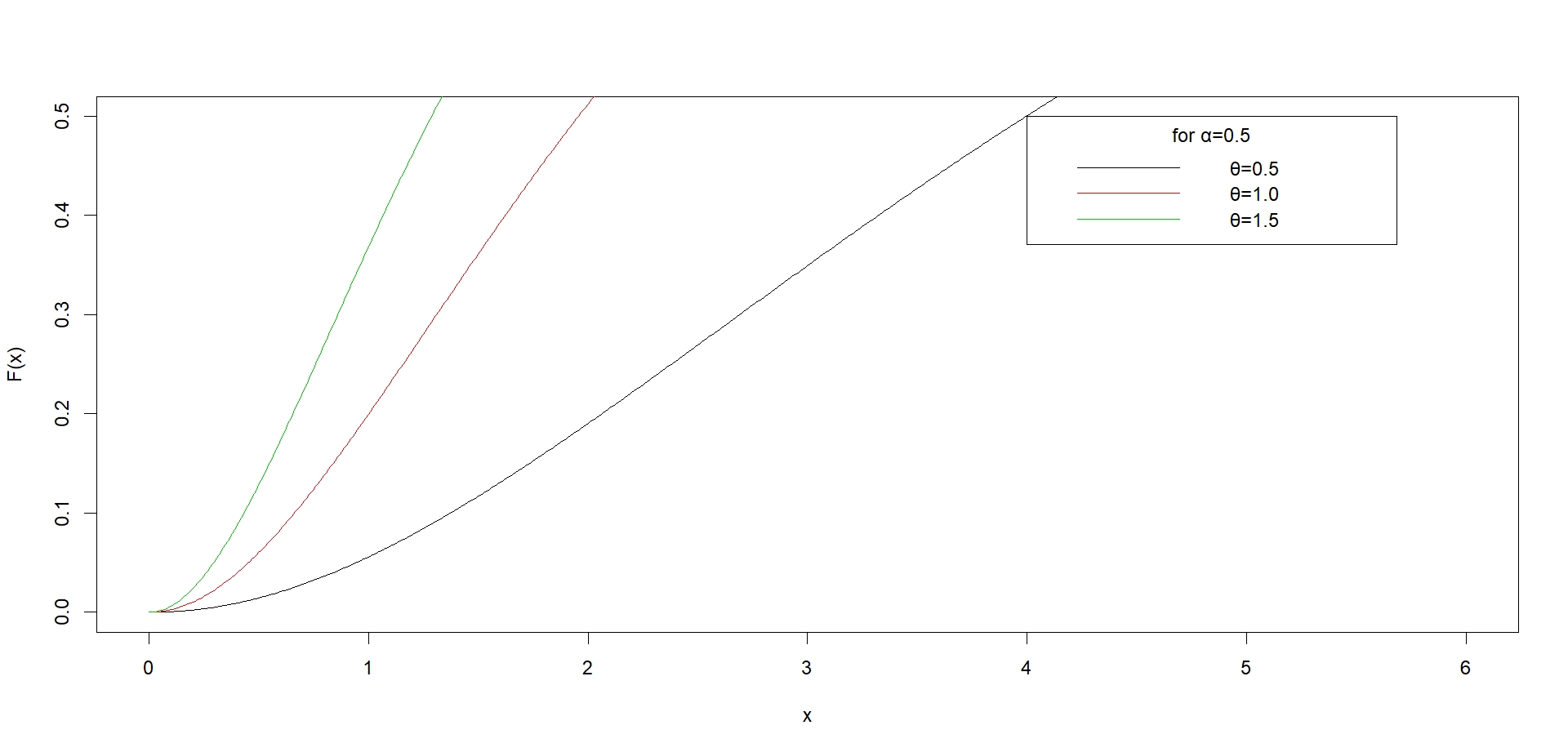}}}%
	\quad
	\subfloat{{\includegraphics[width=6.8cm]{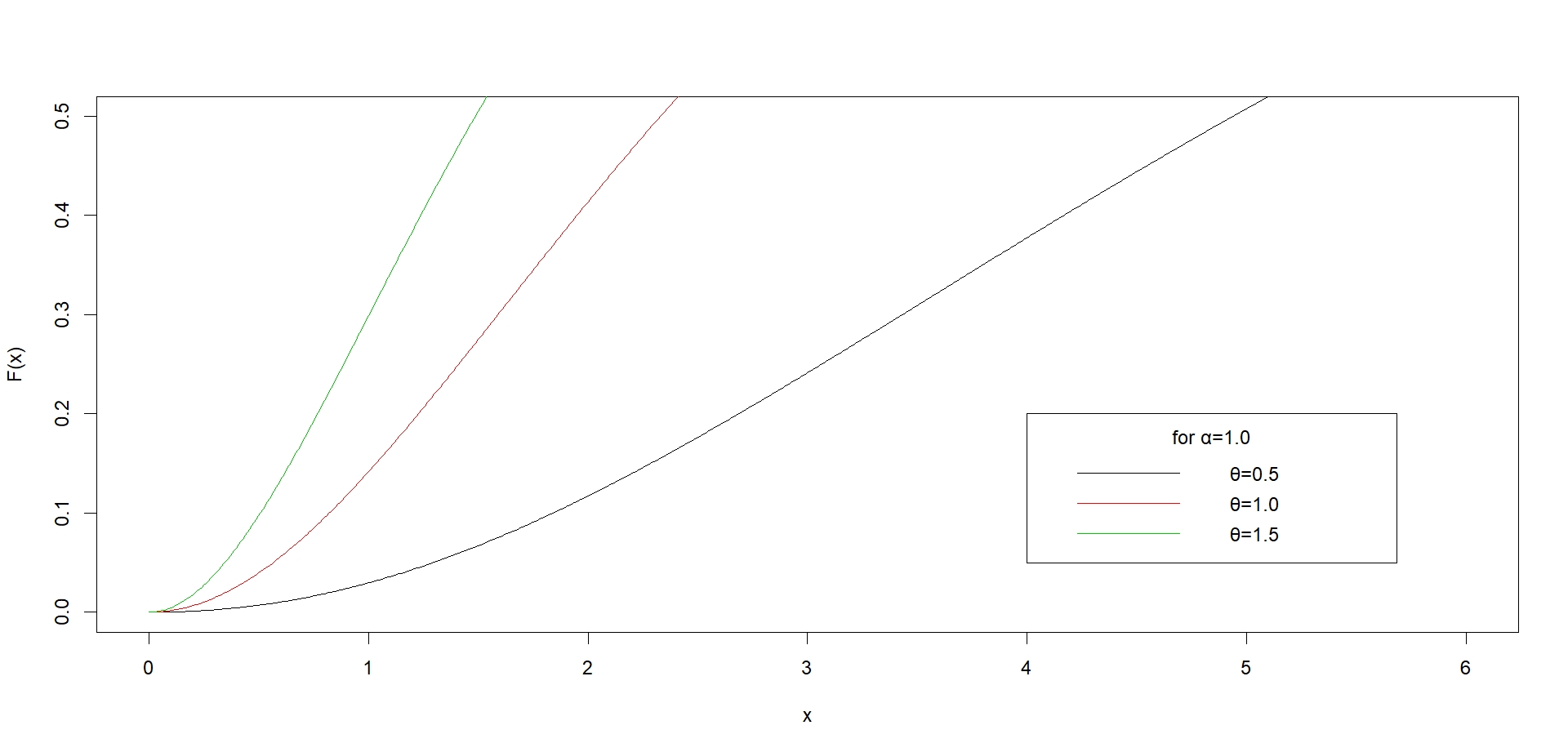}}}%
	\quad
	\subfloat{{\includegraphics[width=6.8cm]{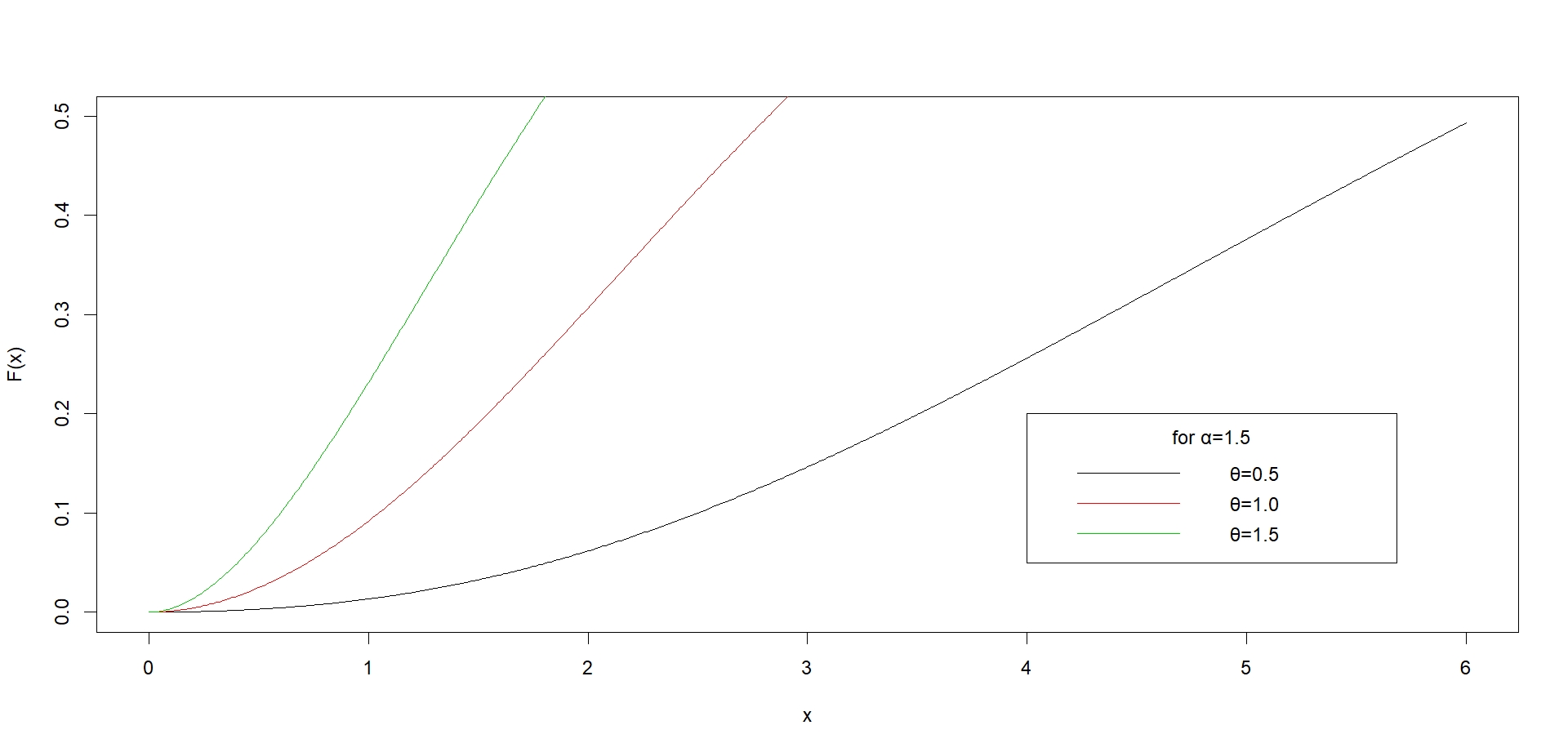}}}%
	\quad
	\subfloat{{\includegraphics[width=6.8cm]{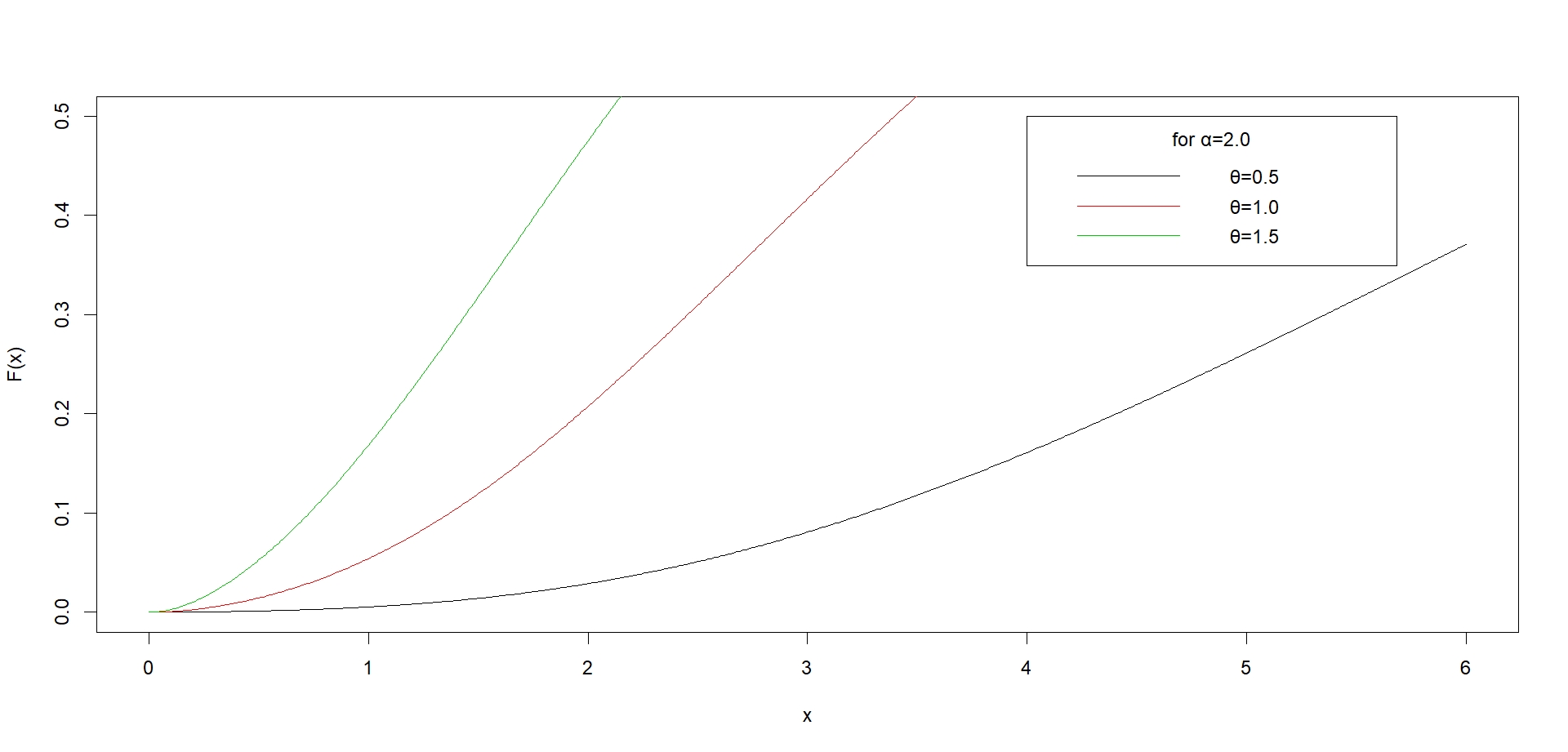}}}%
	\caption{cdf plot of SSD for different values of $\alpha$ and $\theta$}
	\label{fig:cdf}
\end{figure}

\noindent In this study a new more flexible generalized distribution named SSD distribution is proposed which is the mixture of gamma (2, $\theta$) i.e. length biased exponential ($\theta$) with gamma ($\alpha$+2, $\theta$). The pdf of two parameter proposed generalised Lindley distribution with parameters ($\alpha$+2, $\theta$) can be written as
\begin{align}
	f(x,\theta)=\frac{\theta^{\alpha+2}}{\theta^\alpha+(\alpha+1)!} e^{-\theta x} {(x+x^{\alpha+1})}  x>0, \theta>0, \alpha N
\end{align}	

\noindent we have

\[f \left(x \right)=p.f_1+(1-p).f_2\]  

\noindent where
\begin{align}
	p=\frac{\theta^{\alpha+2}}{\theta^\alpha+(1+\theta)!},& \quad f_{1} \left(x\right)=\frac{\theta ^{2} }{\Gamma 2} e^{-\theta x} x^{2-1} =\theta ^{2} x.e^{-\theta x} \nonumber\\
	f_{2} \left(x\right)=\frac{\theta ^{\alpha+2} }{\Gamma(\alpha+2)}& e^{-\theta x} x^{\alpha+2-1} =\frac{\theta ^{\alpha+2} }{\Gamma {\alpha+2}} x^{\alpha+1} e^{-\theta x}\nonumber
\end{align}
\newline The plot of probability density function of SSD distribution is given in figure (\ref{fig:pdf})
\noindent In the above pdf putting $n$=1,2,3,... we can get series of various distributions. The cumulative distribution function for the proposed distribution can be written as 
\begin{align}
F(x)=\int\limits_{0}^{x}f(t)dt= \frac{\theta ^{\alpha } \left[1-(1+\theta x)e^{-\theta x} \right]+\gamma (\alpha +2,\theta x)}{(\alpha +1)!+\theta ^{\alpha } } 
\end{align}
\noindent Where,$\gamma$ is Lower incomplete gamma function, defined as $\gamma (s,x)=\int\limits_{0}^{x}t^{s-1} e^{-t} dt $\newline
The plot of cumulative distribution function of SSD distribution is given in figure (\ref{fig:cdf}).
\section{Statistical properties and Related measures of SSD}
\subsection{Moments, Moment generating function and Characteristics function}

\noindent The moment and moment generating functions play important roles for analysing any distributions functions. The moment generating function characterizes the distribution function. Although, we could not obtain the moments in explicit forms, they can be obtained as infinite summation of beta functions.

\noindent Moments are important in any statistical analysis, especially in applications. It can be used to study the most important features and characteristics of adistribution (e.g. tendency, dispersion, skewness, and kurtosis).

\noindent The \textit{r}-th moment about origin of the generalised Lindley distribution is given by:
\[\mu '_{r} =E(X^{r})=\int\limits_{0}^{\infty }x^{r} f(x)dx \] 
so 
\begin{align}
\mu'_{r}=\frac{\theta ^{\alpha -r} }{(\alpha +1)!+\theta ^{{}^{\alpha } } } \left[(r+1)!+\frac{(\alpha +r+1)!}{\theta ^{\alpha}} \right]
\end{align}
Now putting $r=1, 2, 3{\dots}$ we get the first four moments aspects       \newline
\[\mu '_{1} =E(X)=\frac{2\theta ^{\alpha}+(\alpha +2)! }{\theta(\theta ^{\alpha }+(\alpha +1)!) }\]
\[\mu '_{2} =E(X^2)=\frac{6\theta ^{\alpha}+(\alpha +3)! }{\theta^{2}(\theta ^{\alpha }+(\alpha +1)!) }\]
\[\mu '_{3} =E(X^3)=\frac{24\theta ^{\alpha}+(\alpha +4)! }{\theta^{3}(\theta ^{\alpha }+(\alpha +1)!) }\]
\[\mu '_{4} =E(X^4)=\frac{120\theta ^{\alpha}+(\alpha +5)! }{\theta^{4}(\theta ^{\alpha }+(\alpha +1)!) }\]

\noindent So,
\begin{align}
V(x)=\mu'_{2}-(\mu'_{1})^{2}=\frac{1}{\theta^{2} \left[\theta ^{{}^{\alpha } } +(\alpha +1)!\right]} \left[6\theta ^{\alpha } +(\alpha +3)!-\frac{\left[2\theta ^{\alpha } +(\alpha +2)!\right]^{2} }{\theta ^{{}^{\alpha } } +(\alpha +1)!} \right]
\end{align}

\noindent Now, we derive the moment generating function and characteristic function. The moment generating function is given by the relation
\[M_x(t)=\int\limits_{0}^{\infty} e^{tx}f(x)dx\]
\noindent then, we have,
\[M_x (t)=\int\limits_{0}^{\infty} e^{tx}\frac{\theta^{\alpha+2}}{\theta^\alpha+(\alpha+1)!} e^{-\theta x} {(x+x^{\alpha+1})}dx\]
After solving the above equation, we get,
\begin{align}
M_x (t)=\frac{\theta^{\alpha+2}}{(\theta-t)^{\alpha+2}} \frac{(\theta-t)^{\alpha+2}+(\alpha+1)!}{\theta^{\alpha}+(\alpha+1)!} 
\end{align}
Now we replace $ it $ for $ t $ in equation number (2.3) we get the corresponding Characteristics function for the SSD distribution can be written as in equation number (2.4).
\begin{align}
\phi_x (t)=\frac{\theta^{\alpha+2}}{(\theta-it)^{\alpha+2}} \frac{(\theta-it)^{\alpha+2}+(\alpha+1)!}{\theta^{\alpha}+(\alpha+1)!}
\end{align}

\subsection{Hazard function and Mean Residual Life function}
\begin{figure}[H]
	\centering
	\subfloat{{\includegraphics[width=6.8cm]{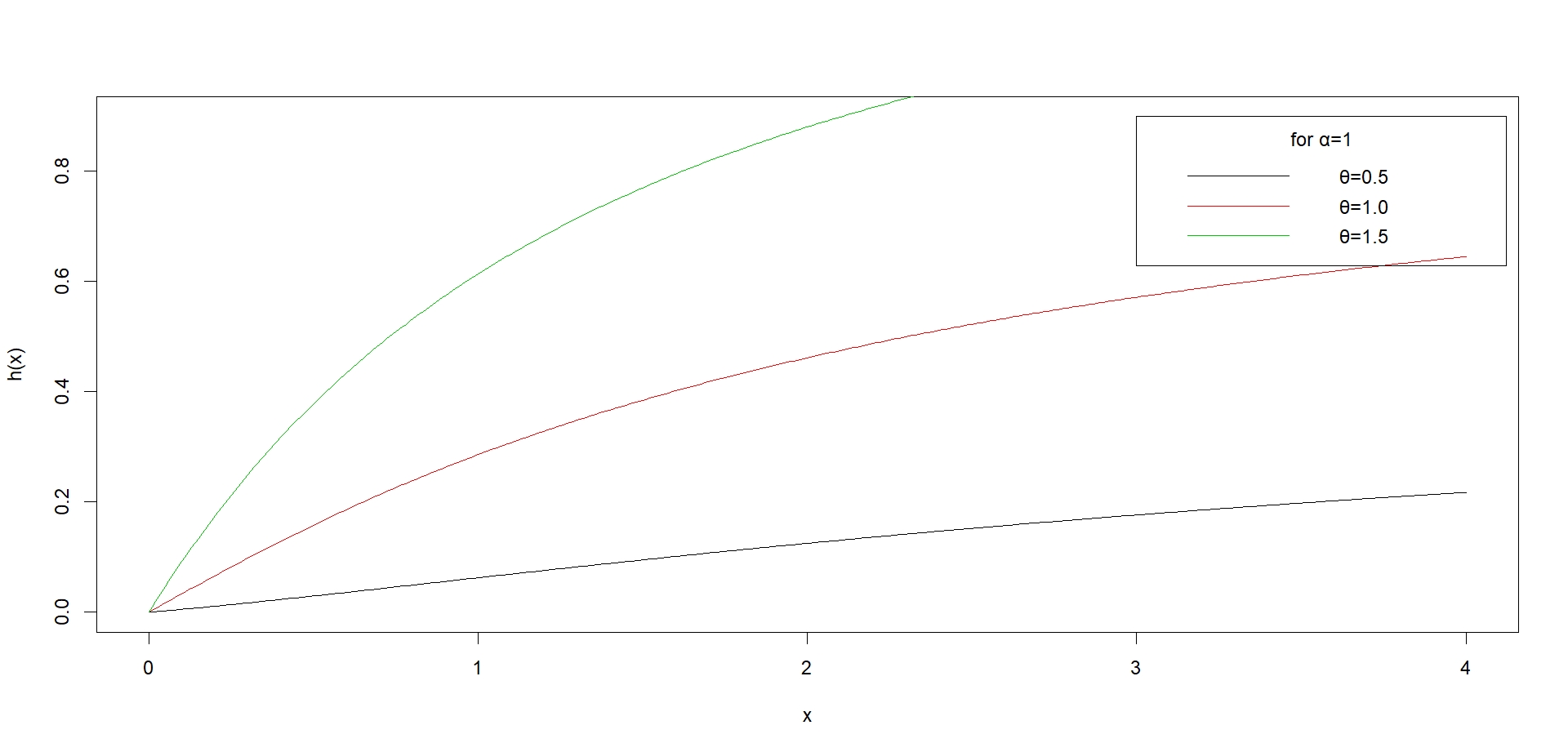}}}%
	\quad
	\subfloat{{\includegraphics[width=6.8cm]{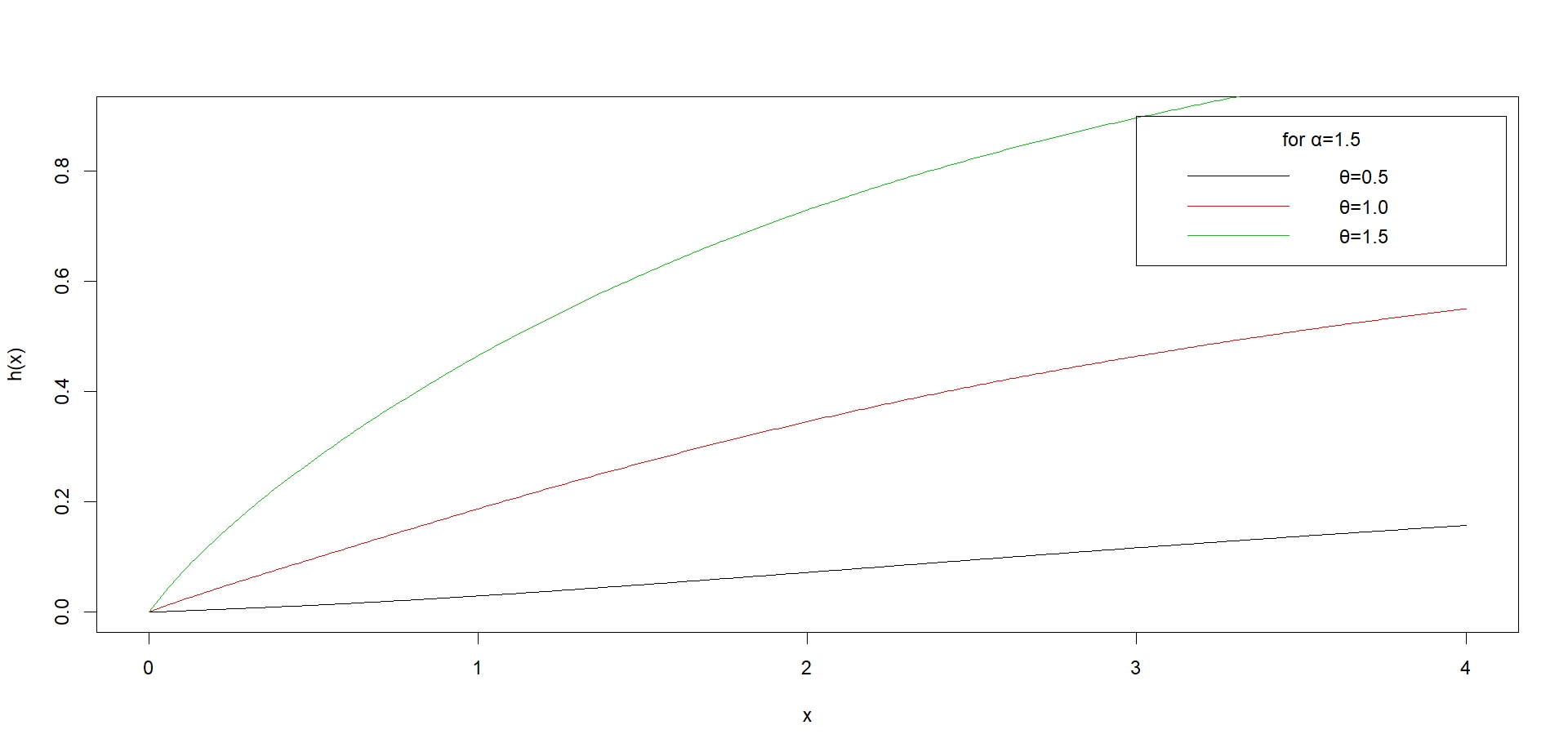}}}%
	\quad
	\subfloat{{\includegraphics[width=6.8cm]{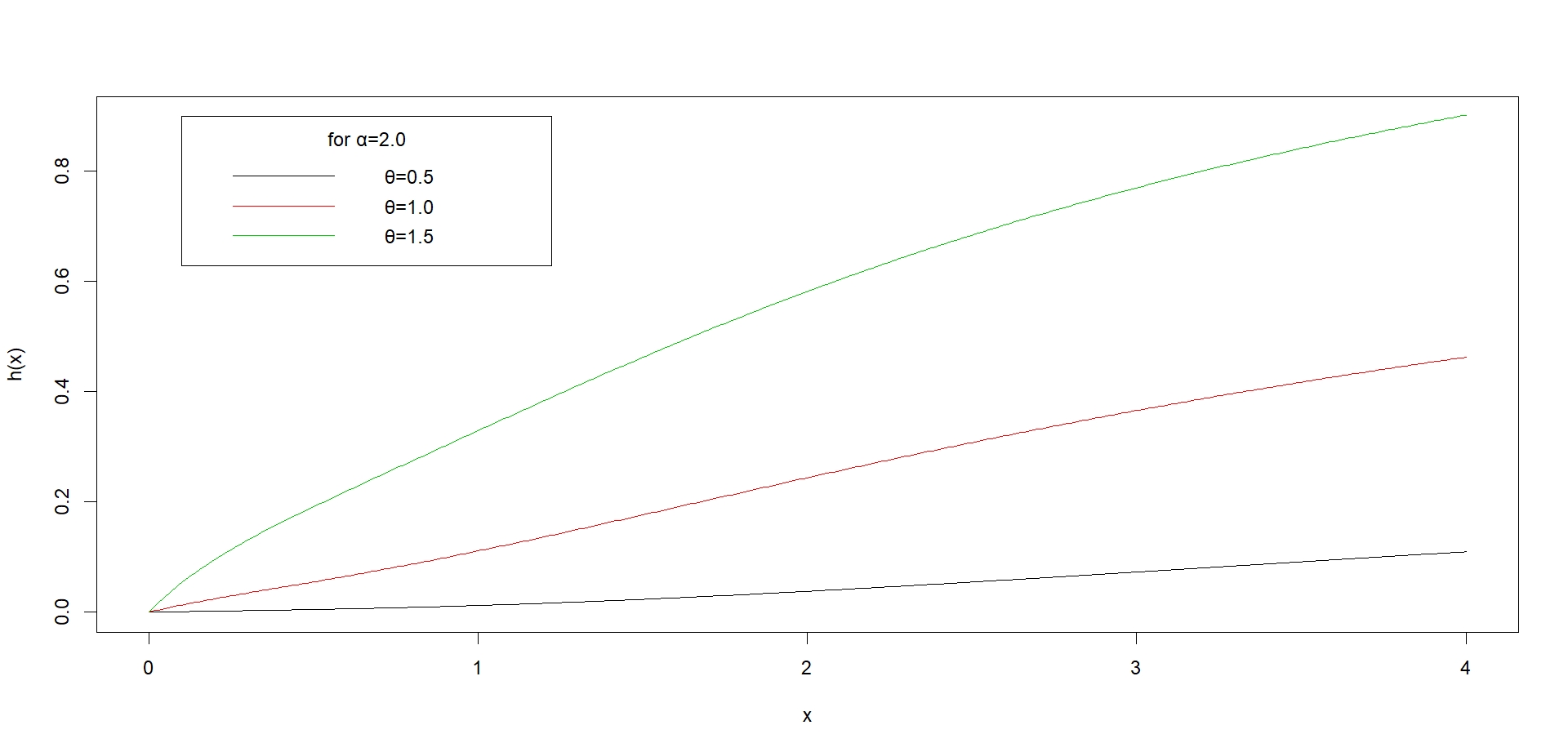}}}%
	\quad
	\subfloat{{\includegraphics[width=6.8cm]{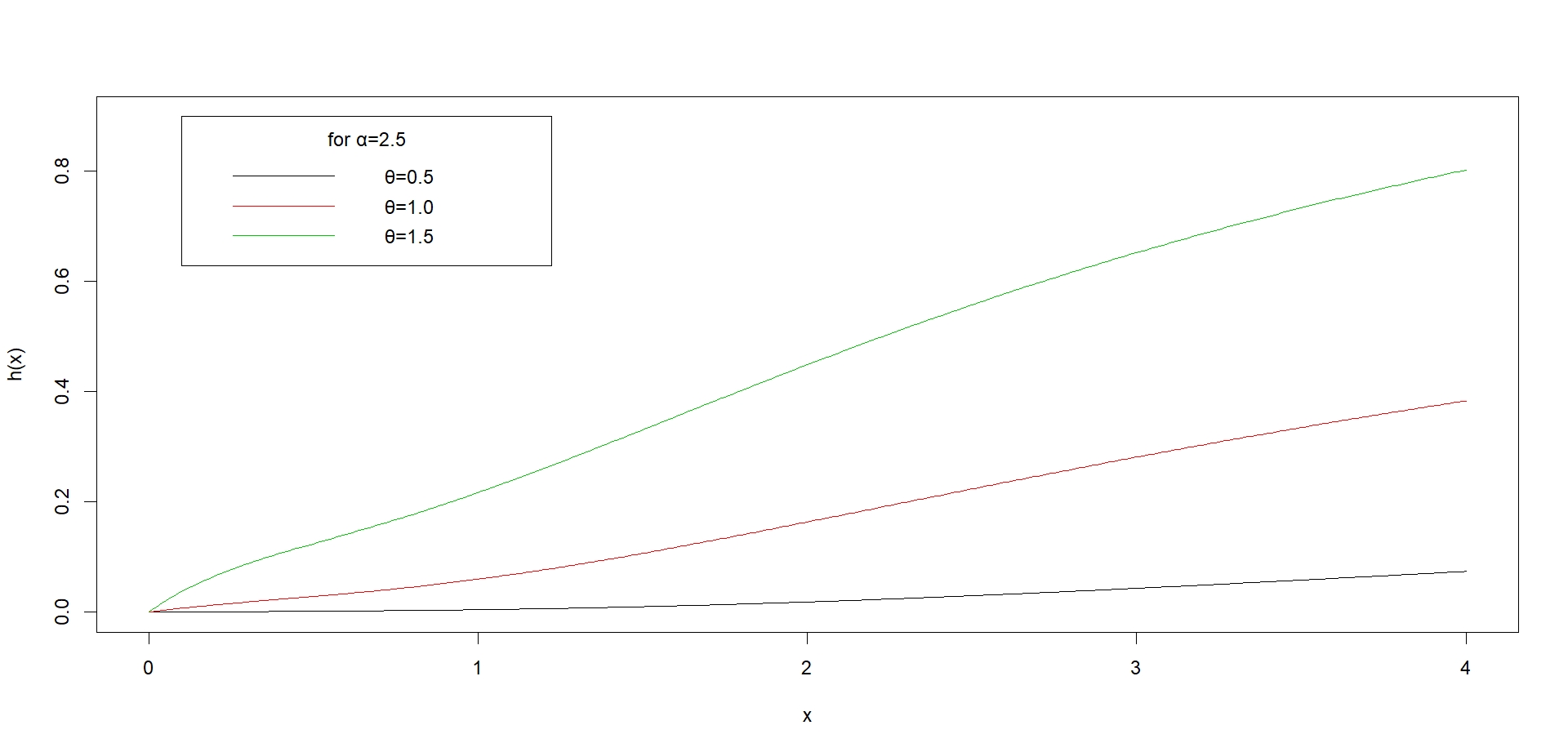}}}%
	\caption{Hazard plot for SSD for different values of $\alpha$ and $\theta$}
	\label{fig:haz}
\end{figure}

\begin{figure}[H]
	\centering
	\subfloat{{\includegraphics[width=6.8cm]{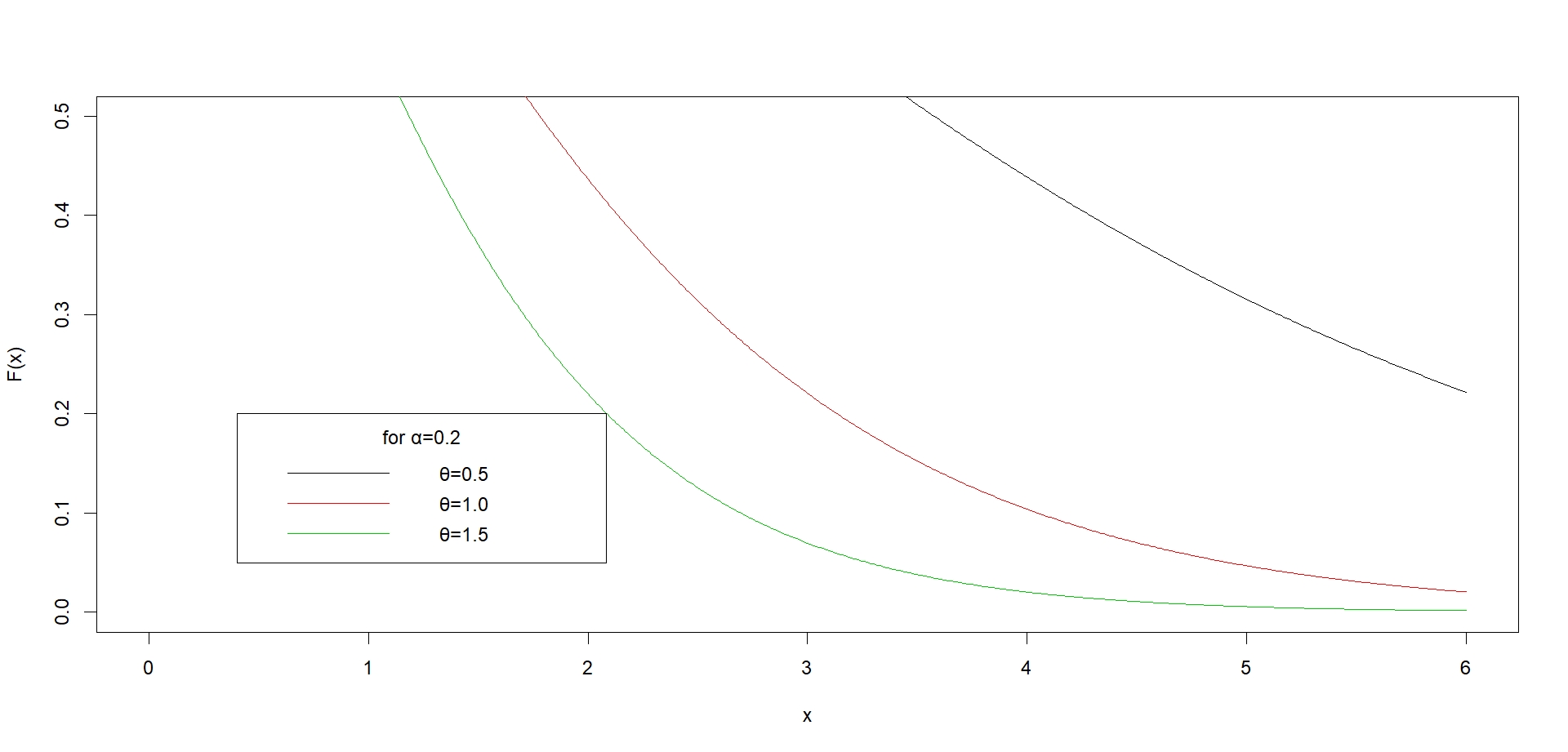}}}%
	\quad
	\subfloat{{\includegraphics[width=6.8cm]{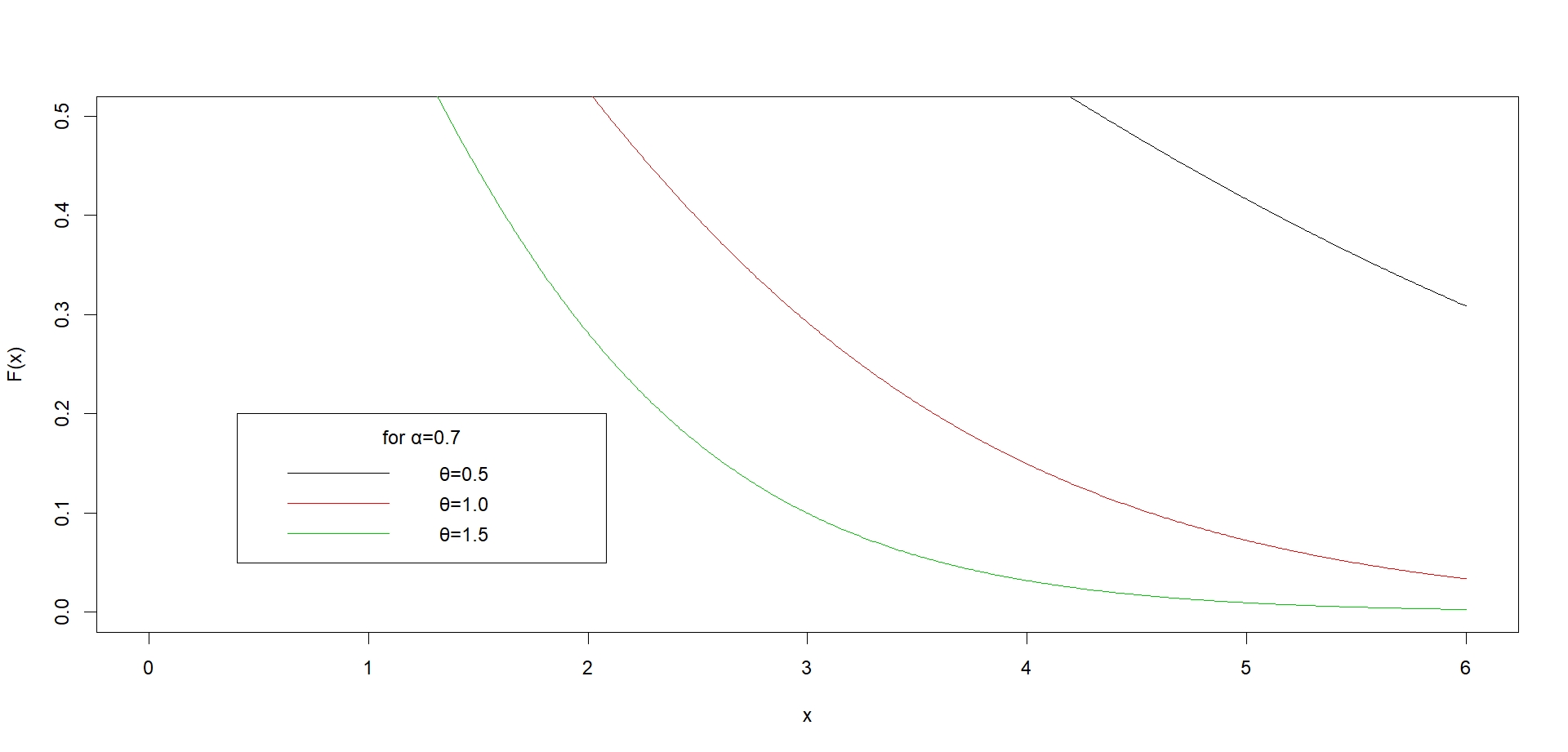}}}%
	\quad
	\subfloat{{\includegraphics[width=6.8cm]{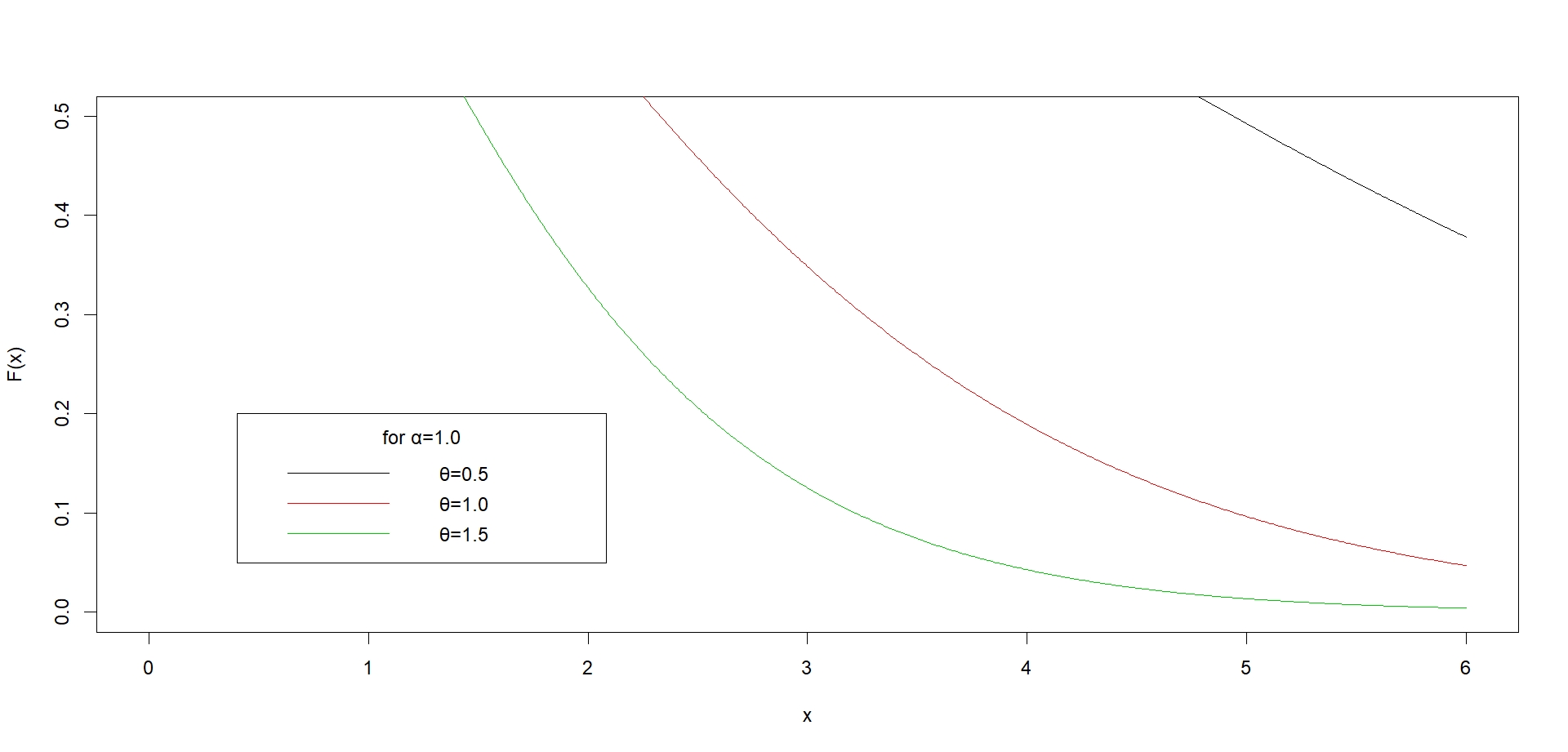}}}%
	\quad
	\subfloat{{\includegraphics[width=6.8cm]{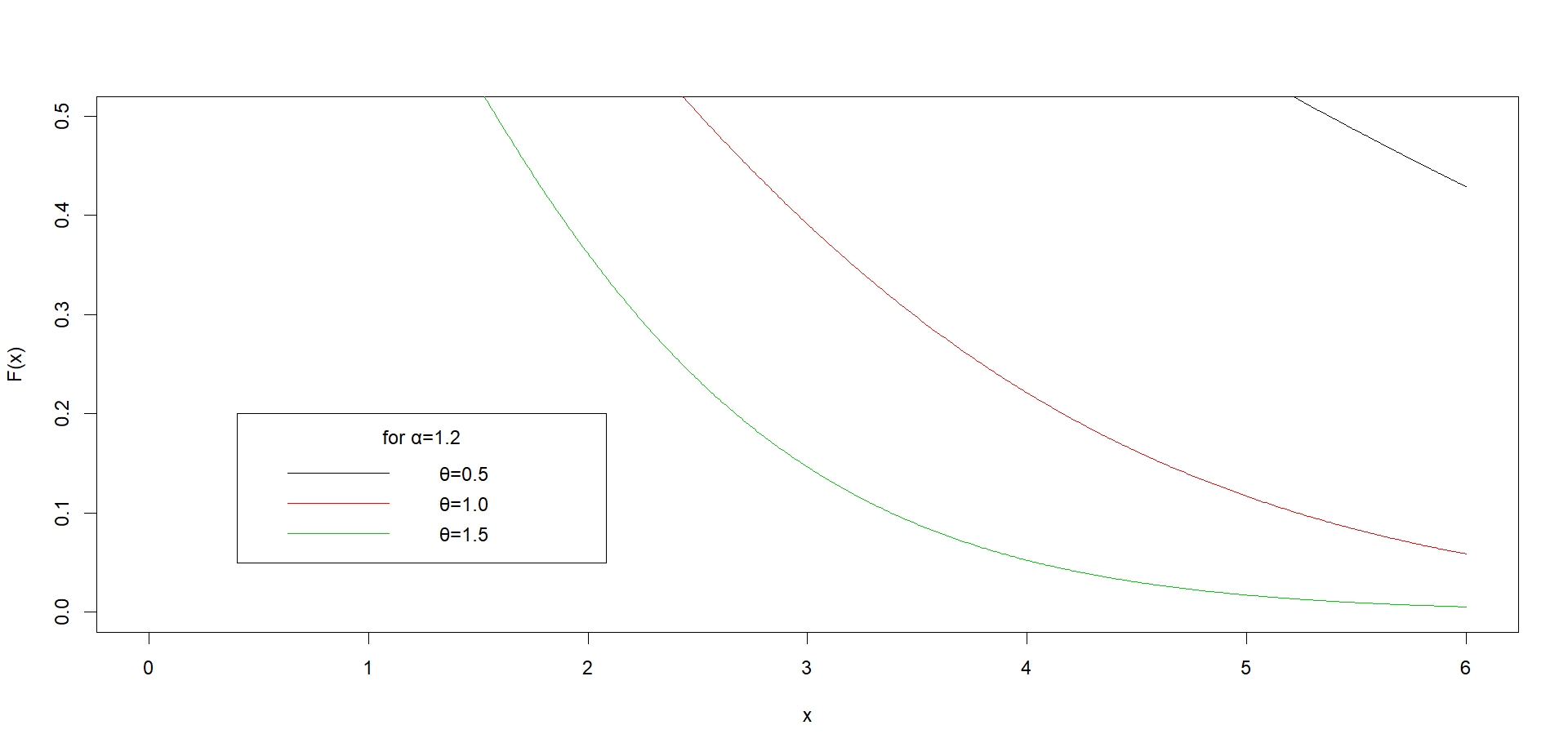}}}%
	\caption{Survival plot for SSD for different values of $\alpha$ and $\theta$}
	\label{fig:sur}
\end{figure}

\noindent Let \textit{f }(\textit{x}) and \textit{F }(\textit{x}) be the p.d.f. and c.d.f. of acontinuous random variable. The hazard rate function (alsoknown as the failure rate function), survival function and the mean residual life function of \textit{X} are respectively given as
\[h(x)=\frac{f(x)}{R(x)}=\frac{f(x)}{1-F(x)}
=\frac{\left(\frac{\theta ^{\alpha +2} }{(\alpha +1)!+\theta ^{\alpha } } \right)e^{-\theta x} \left(x+x^{\alpha +1} \right)}{1-\left(\frac{\theta ^{\alpha } \left\{1-\left(1+\theta x\right)e^{-\theta x} \right\}+\gamma \left(\alpha +2,\theta x\right)}{(\alpha +1)!+\theta ^{\alpha } } \right)} \] 
\begin{align}\label{eq:haz1}
h(x)=\frac{\theta ^{\alpha +2} e^{-\theta x} \left(x+x^{\alpha +1} \right)}{(\alpha +1)!+\theta ^{\alpha } \left(1+\theta x\right)e^{-\theta x} -\gamma \left(\alpha +2,\theta x\right)} 
\end{align}
Now differentiating \eqref{eq:haz1} with respect to $x$ we get 
\begin{align}\tiny
	h'(x)=\frac{\theta^{\alpha+2}e^{-\theta x}\left[\left(1+(\alpha+1)x^\alpha-\theta x(1+x^\alpha)\right)\left((\alpha+1)!+\theta^{\alpha}(1+\theta x)e^{-\theta x}-\gamma(\alpha+2,\theta x)\right)\atop
		+(x+x^{\alpha+1})(\theta^{\alpha+2}xe^{-\theta x}+\gamma'(\alpha+2,\theta x))\right]}{\left[(\alpha+1)!+\theta^\alpha(1+\theta x)e^{-\theta x}-\gamma(\alpha+2,\theta x)\right]^2}
\end{align}
Now as $x\to{0}$ we get
\begin{align}\label{eq:haz2}
	h'(0)=\frac{\theta^{\alpha+2}}{(\alpha+1)!+\theta^{\alpha}}>0 \forall \quad \theta>0,\alpha>0
\end{align}
Hence from \eqref{eq:haz2} we can say that hazard is increasing and the plot of hazard function of SSD is given in figure (\ref{fig:haz}).
\noindent The survival function of SSD is given as
\begin{align}
	S(x)=\int\limits_{x}^{\infty}f(t)dt=1-F(x)= \frac{(\alpha+1)!+\theta ^{\alpha }(1+\theta x)e^{-\theta x}-\gamma (\alpha +2,\theta x)}{(\alpha +1)!+\theta ^{\alpha } } 
\end{align}
The plot of survival function of SSD is given in figure (\ref{fig:sur})
\noindent Mean residual life function for the SSD distribution can be given as
\[m(x)=E(X-x|X>x)=\frac{1}{S(x)} \int\limits_{x}^{\infty} tf(t) dt-x\]           
\[=\frac{1}{S(x)} \int\limits_{x}^{\infty }t \left(\frac{\theta ^{\alpha +2} }{(\alpha +1)!+\theta ^{\alpha } } \right)e^{-\theta t} \left(t+t^{\alpha +1} \right)-x\] 
\[=\frac{1}{S(x)} \left(\frac{\theta ^{\alpha +2} }{(\alpha +1)!+\theta ^{\alpha } } \right)\int _{x}^{\infty }t e^{-\theta t} \left(t+t^{\alpha +1} \right)-x\] 
\[=\frac{1}{S(x)} \left(\frac{\theta ^{\alpha +2} }{(\alpha +1)!+\theta ^{\alpha } } \right)\left[\frac{\left(\theta ^{2} x^{2} +2\theta x+2\right)e^{-\theta x} }{\theta ^{3} } +\frac{\Gamma \left(\alpha +3,\theta x\right)}{\theta ^{\alpha +3} } \right]-x\] 
After solving the above integral, we get,
\begin{align}
m(x)=\frac{\left[\theta ^{\alpha } \left(\theta ^{2} x^{2} +2\theta x+2\right)e^{-\theta x} +\Gamma\left(\alpha +3,\theta x\right) \right]}{\theta \left[(\alpha +1)!+\theta ^{\alpha } (1+\theta x)e^{-\theta x} -\gamma \left(\alpha +2,\theta x\right)\right]} -x
\end{align}
Where $  \ {\Gamma {(*,*)}}   $  is upper incomplete gamma function , $ \ {\Gamma {(s,x)}} =\int _{x}^{\infty }t^{s-1} e^{-t} dt $  and ${\mathop{\lim }\limits_{x\to 0}}  \ {\Gamma{(s,x)}} =(s-1)!$ and $\gamma(s,x)=\int _{0}^{x }t^{s-1} e^{-t} dt$ is the lower incomplete gamma function.

\subsection{ R\'{e}nyi  entropy}
\noindent Entropy is a measure of the uncertainty associated with a random variable. The Shannon entropy is a measure of the average information contentone is missing when one doesn't know the value of the random variable. A usefulgeneralization of Shannon entropy is the R\'{e}nyi entropy. It is also important in quantum information, where it can beused as a measure of entanglement; e.g. see Shannon (1948); Renyi (1961).

\noindent The R\'{e}nyi entropy of the proposed distribution is given as

\[T_R (\gamma)=\frac{1}{1-\gamma}\log\int {f^{\gamma}(x)dx}\]
\[T_R (\gamma)=\frac{1}{1-\gamma}\log\int {\left [\frac{\theta^{\alpha+2}}{\theta^\alpha+(\alpha+1)!} e^{-\theta x} {(x+x^{\alpha+1})}\right]^{\gamma}dx}\]
\noindent hence after simplification above integral we get,
\begin{align}
T_R (\gamma)=\frac{1}{1-\gamma}\log\left [\frac{\theta^{\gamma(\alpha+2)}}{\{ {\theta^{\alpha}+(\alpha+1)!} \}^{\gamma}}\sum_{k=0}^{\gamma }\binom{\gamma }{k}\frac {(\alpha k+\gamma )!}{(\theta \gamma)^{\alpha k+\gamma +1} } \right ]
\end{align}

\subsection{Bonferroni and Lorenz curves}
\noindent The Bonferroni and Lorenz curves (Bonferroni, 1930) and Bonferroni and Gini indices have applications not only in economics to study income and poverty, but also in other fields like reliability, demography, insurance and medicine. The Bonferroni curve $ B(p) $and Lorenz curves $ L(p) $are defined as

\[B(p)=\frac{1}{p\mu } \int _{0}^{q}xf(x)dx =\frac{1}{p\mu} \left[\int_{0}^{\infty}xf(x)dx-\int_{q}^{\infty}xf(x)dx\right]=\frac{1}{p\mu }\left[\mu-\int_{q}^{\infty}xf(x)dx\right]\] 
and
\[L(p)=\frac{1}{p } \int _{0}^{q}xf(x)dx =\frac{1}{p} \left[\int_{0}^{\infty}xf(x)dx-\int_{q}^{\infty}xf(x)dx\right]=\frac{1}{p\mu }\left[\mu-\int_{q}^{\infty}xf(x)dx\right] \] 

\noindent Using the proposed pdf, we get 
\begin{align}
	\int_{q}^{\infty}xf(x)dx=\left(\frac{\theta^{\alpha+2}}{(\alpha +1)!+\theta^{\alpha}}\right)\left[\frac{\left(\theta^2q^2+2\theta q+2\right)e^{-\theta q}}{\theta^3}+\frac{\Gamma{(\alpha+3,\theta q)}}{\theta^{\alpha+3}}\right]
\end{align}
So from the above equations we get
\begin{align}
B(p)=\frac{1}{p}\left[1-\frac{\left\{\theta^{\alpha}\left(\theta ^{2} q^{2} +2\theta q+2\right)e^{-\theta q}+\Gamma{(\alpha+3,\theta q)}\right\}}{2\theta^{\alpha}+(\alpha +2)!}\right]
\end{align}
and
\begin{align}
L(p)=1-\frac{\left\{\theta^{\alpha}\left(\theta ^{2} q^{2} +2\theta q+2\right)e^{-\theta q}+\Gamma{(\alpha+3,\theta q)}\right\}}{2\theta^{\alpha}+(\alpha +2)!}
\end{align}

\subsection{ Order Statistics} 

\noindent Let $ x_1, x_2,\dots x_n $ be a random sample of size \textit{n }from the SSD distribution. Let $ X_{(1)}<X_{(2)}<\dots<X_{(n)} $ denote the corresponding order statistics. The p.d.f. and the c.d.f. of the \textit{k }th order statistic, say $ Y=X_{(k)} $ are given by

\[ {f_{Y} (y)=\frac{n!}{(k-1)!(n-k)!} F^{k-1}(y)\{1-F(y)\}^{n-k}f(y)}\] 
or
\begin{align}
f_{Y}(y)=\frac{n!}{(k-1)!(n-k)!}\sum_{l=0}^{n-k}\binom{n-k}{l}(-1)^{l}F^{k+l-1}(y)f(y)
\end{align}
and 
\[F_{Y}(y)=\sum_{j=k}^{n}\binom{n}{j}F^{j}(y)\{1-F(y)\}^{n-j}\]
or
\begin{align}
F_{Y}(y)=\sum_{i=k}^{n}\sum_{l=0}^{n-j}\binom{n}{j}\binom{n-i}{l}(-1)^{l}F^{i+l}(y)
\end{align}

\noindent Now, using equation number (1.2) and (1.3) in equation (2.10) and (2.11) we get the corresponding pdf and the cdf of $ k-th $ order statistics of the SSD distribution are obtained as
\[\begin{array}{l} {f_{Y} (y)=\frac{n!}{(k-1)!(n-k)!} \sum \limits_{l=0}^{n-k}\left(\begin{array}{l} {n-k} \\ {\, \, \, \, \, l} \end{array}\right) (-1)^{l} \left(\frac{\theta ^{\alpha } }{(\alpha +1)!+\theta ^{\alpha } } \right)^{k+l} } \\ {\, \, \, \, \, \, \, \, \, \, \, \, \, \, \, \, \, \, \, \, \left[\left\{1-\left(1+\theta x\right)e^{-\theta x} \right\}+\frac{\gamma \left(\alpha +2,\theta x\right)}{\theta ^{\alpha } } \right]^{k+l-1} \left(x+x^{\alpha +1} \right)e^{-\theta x} } \end{array}\] 
And
\[F_{Y} (y)=\sum _{i=k}^{n}\sum _{l=0}^{n-i}\left(\begin{array}{l} {n} \\ {i} \end{array}\right) \left(\begin{array}{l} {n-i} \\ {\, \, \, \, \, l} \end{array}\right) (-1)^{l} \left[\frac{\theta ^{\alpha } }{(\alpha +1)!+\theta ^{\alpha } } \right]^{i+l} \left[\left\{1-\left(1+\theta x\right)e^{-\theta x} \right\}+\frac{\gamma \left(\alpha +2,\theta x\right)}{\theta ^{\alpha } } \right]^{i+l} \] 

\subsection{Total time on test (TTT)}
We know that distribution function $F(t)$ and and $\mu$ the mean time to failure (MTTF). The $F(t)$ is continuous and strictly increasing which indicates that $F^{-1}(t)$ exists, thus the Total time on test (TTT) of $F(t)$ is defined as 
\begin{align}
	\phi(x)=\frac{1}{\mu}\int\limits_{0}^{F^{-1}(x)}\left[1-F(t)\right]dt \quad\text{for}\quad 0\leq\mu\leq1
\end{align}
The value of $\phi(x)$ is interpreted as the area below $\frac{1-F(t)}{\mu}$  between 0 and $F^{-1}(\mu)$.
The TTT plot, an empirical and scale independent plot based on failure data and corresponding to asymptotic curve, The scaled TTT-transformation see Barlow and Campo (1975) is used to illustrate some test statistics for testing exponentiality.
\begin{figure}[H]
	\centering
	\subfloat[TTT plot for 1st data set of SSD]{{\includegraphics[width=6.8cm]{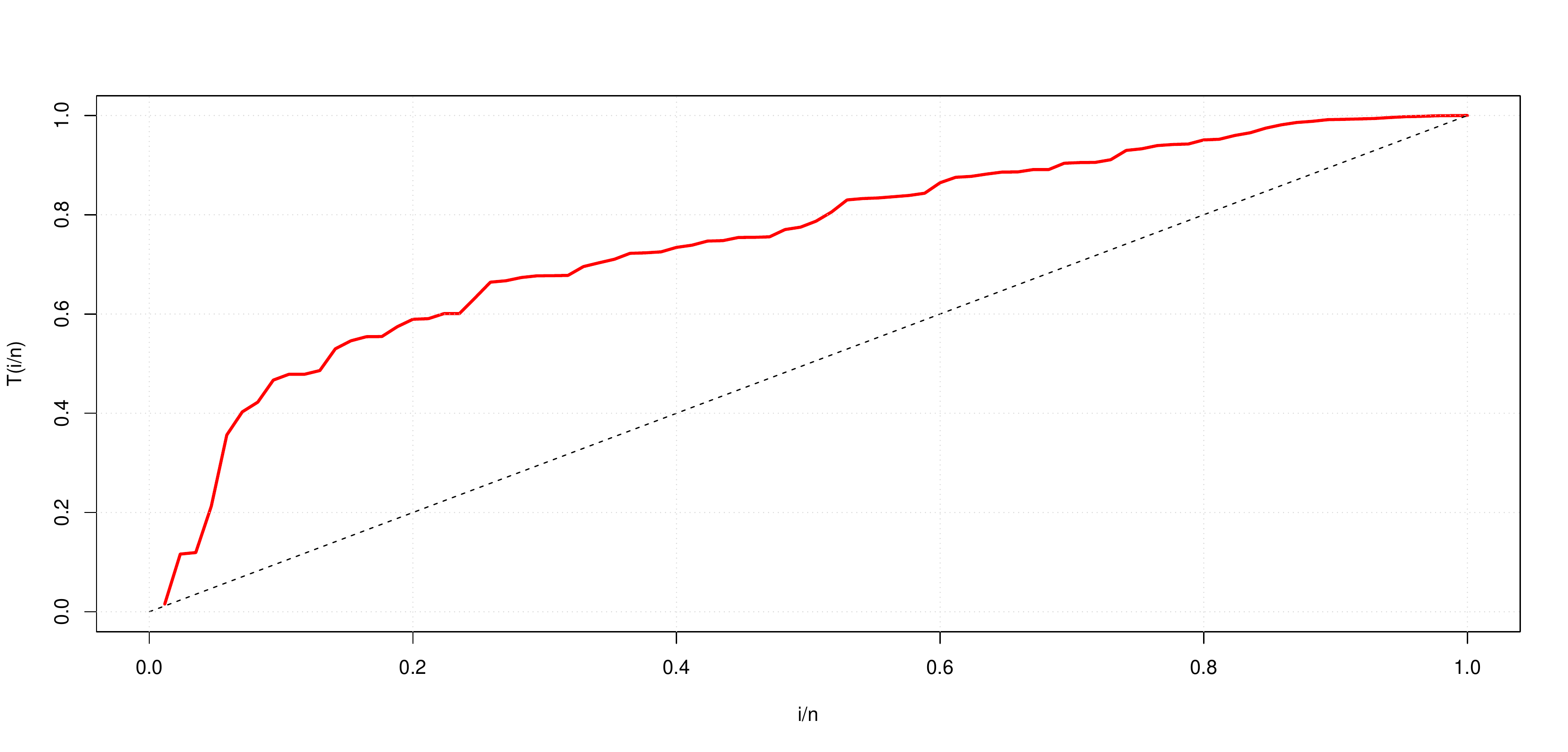}}}%
	\quad
	\subfloat[TTT plot for 2nd data set of SSD]{{\includegraphics[width=6.8cm]{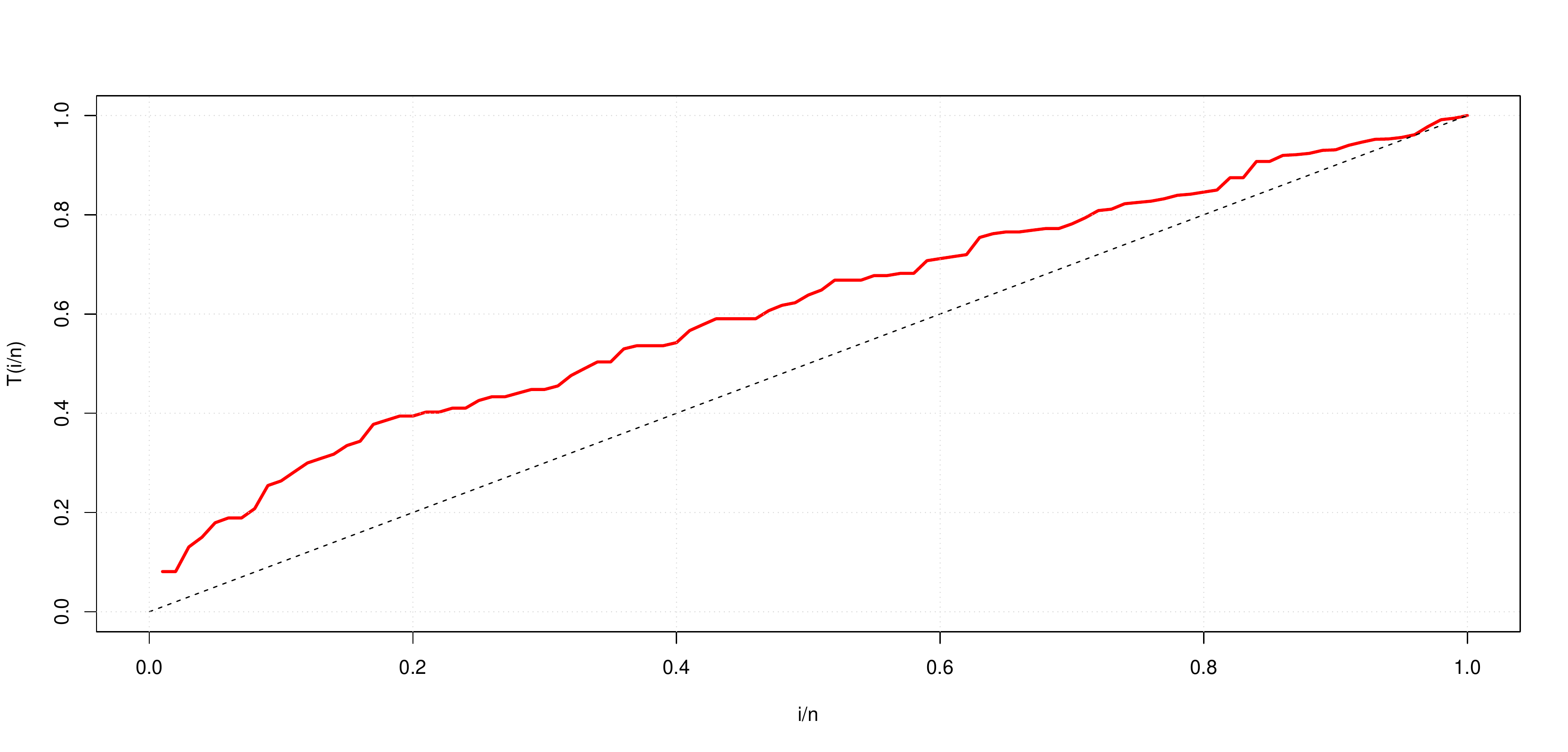} }}%
	\caption{TTT plot for fitted data sets}
	\label{fig:ttt}
\end{figure}

\section{ Estimation of the parameter}
\noindent Above discussed generalisation of Lindley distribution is of two parameters which are estimated by method of maximum likelihood. The likelihood function for the proposed distribution can be written as 

\[L(\theta )=\prod_{i=1}^{n}{\left [ \frac{\theta ^{\alpha +2}}{\theta ^{\alpha }+(\alpha +1)!} \right ]}e^{-\theta x_{i}}(x_{i}+x_{i}^{\alpha +1})\]

\noindent or 
\[L(\theta )={\left [ \frac{\theta ^{n(\alpha +2)}}{\{\theta ^{\alpha }+(\alpha +1)!\}^{n}} \right ]}e^{-\theta \sum_{i=1}^{n}x_{i}}\prod_{i=1}^{n}(x_{i}+x_{i}^{\alpha +1})\]

\noindent Now, log-likelihood can be given as
\[ \log L(\theta )=n(\alpha +2)\log\theta-n\log\{\theta ^{\alpha }+(\alpha +1)!\}-\theta \sum_{i=1}^{n}x_{i}+ \sum_{i=1}^{n}\log(x_{i}+x_{i}^{\alpha +1}) \]

\noindent Differentiating the above equation with respect to $\theta $ and $\alpha $ partially, we get,
\begin{align}\label{eq:mle1}
\frac{\partial \log L}{\partial \theta}=\frac{n(\alpha +2)}{\theta }-\frac{n\alpha \theta ^{\alpha -1}}{\theta ^{\alpha }+(\alpha +1)!}-\sum_{i=1}^{n}x_{i}
\end{align}
and
\begin{align}\label{eq:mle2}
\frac{\partial \log L}{\partial \alpha }=n\log\theta-\frac{n \left [\frac{\partial (\alpha +1)!}{\partial \alpha } +\theta^{\alpha}\log{\theta} \right ]}{\theta ^{\alpha }+(\alpha +1)! }+\frac{\partial }{\partial \alpha }\left [  \sum_{i=1}^{n}\log (x_{i}+x_{i}^{\alpha +1})\right ]
\end{align}
\noindent Which are non-linear equations and cannot be solved analytically so applying numerical method, we use Newton-Raphson algorithm to evaluate the equations (\ref{eq:mle1}) and (\ref{eq:mle2}).

\section{ Application on Real Data}

The applications of SSD distribution have been discussed with the following two data sets. First data set is relating to failure times of mechanical components reported in the book "Weibull Models" by Murthy et al. (2004) page number-297 and the second data set is waiting time (measured in min) bank customers before service is being rendered, this data set previously used by Ghitany et al. (2008). The graph of total test time (TTT) is shown in the figure (\ref{fig:ttt}).

For the above two dataset, SSD has been fitted along with two- parameter distributions including Shukla distribution proposed by Shukla et al. (2019), gamma distribution, RKD and one parameter lifetime distributions including exponential, Lindley, length biased exponential distribution (LBED) used in Singh and Das (2020). The ML estimates, value of $-$2log \textit{L}, Akaike Information criteria (AIC), Corrected Akaike Information criteria (AICc), K-S statistics and p-value of the fitted distributions are presented in tables 2 and 3.The AIC, BIC, AICc and K-S Statistics are computed using the following formulae:
\begin{align}
	AIC&=-2loglik+2k,\qquad BIC=-2loglik+k\log n\nonumber\\
	AICc&=AIC+\frac{2k^2+2k}{n-k-1},\qquad  D=\sup\limits_{x}|F_{n}{(x)}-F_{0}{(x)}|\nonumber
\end{align}
where $k$= the number of parameters, $n$= the sample size, and the $F_{n}{(x)}$=empirical distribution function and $ F_{0}{(x)}= $   is the theoretical cumulativedistribution function. The best distribution is the distribution corresponding to lower values of$-$2log \textit{L}, AIC, BIC, AICc and K-S statistics and higher p-value.\newline

\begin{table}[H]
	\centering
	\caption{MLE's, - 2ln L, AIC, K-S and p-values of the fitted distributions for the first dataset.}\label{tb:data1}
	\tabcolsep=7.5pt
	\begin{tabular}{ccccccccc}
		\hline \hline
		\multirow{2}*{Distribution} & \multicolumn{2}{c}{Estimate} & \multirow{2}*{-2LL} & \multirow{2}*{AIC} & \multirow{2}*{BIC} & \multirow{2}*{AICc} & \multirow{2}*{K-S} & \multirow{2}*{p-value}\\\cline{2-3}
		& $\alpha$& $\theta$& & & & & &\\\hline
		SSD & 5.7104 & 2.7713  & 258.73 & 262.73 & 267.62 & 262.88 & 0.0808 &  0.6354\\
		SD & 5.1116 & 1.8652 & 285.35 & 289.35 & 294.23 & 289.49 & 0.1451 & 0.0305\\
		RKD & 6.6402 & 2.5467 & 270.66 & 274.66 & 279.55 & 274.81 & 0.0982 & 0.2557\\
		Gamma & 1.3769 & 3.5285 & 280.79 & 285.67 & 280.94 & 276.79 &  0.0919 & 0.3201\\
		LBED &  & 0.7805 & 289.52 & 291.52 & 293.97 & 291.57 &  0.1686 & 0.0079\\
		Lindley &  & 0.6297 & 307.93 & 309.93 & 312.38 & 309.98 &  0.2304 & 0.0000\\
		Exponential &  & 0.3902 & 329.98 & 331.98 & 334.42 & 332.02 & 0.2914 & 0.0000\\
		\hline \hline	
	\end{tabular}
\end{table}

\begin{table}[H]
	\centering
	\caption{MLE's, - 2ln L, AIC, K-S and p-values of the fitted distributions for the second dataset.}\label{tb:data1}
	\tabcolsep=7.5pt
	\begin{tabular}{ccccccccc}
		\hline \hline
		\multirow{2}*{Distribution} & \multicolumn{2}{c}{Estimate} & \multirow{2}*{-2LL} & \multirow{2}*{AIC} & \multirow{2}*{BIC} & \multirow{2}*{AICc} & \multirow{2}*{K-S} & \multirow{2}*{p-value}\\\cline{2-3}
		& $\alpha$& $\theta$& & & & & &\\\hline
		SSD & 0.0143 & 0.2032  & 634.60 & 638.60 & 643.81 & 638.72 & 0.0425 &  0.9937\\
		SD & 1.0848 & 0.2070 & 635.17 & 639.17 & 644.38 & 639.29 & 0.0450 & 0.9874\\
		RKD & 6.6402 & 2.5467 & 636.73 & 640.73 & 645.94 & 640.85 & 0.0504 & 0.9616\\
		Gamma & 0.2034 & 2.0088 & 634.60 & 638.60 & 643.81 & 638.72 & 0.0425 & 0.9936\\
		LBED &  & 0.2025 & 634.60 & 636.60 & 639.21 &  636.64 &  0.0427 & 0.9922\\
		Lindley &  & 0.1866 & 638.07 & 640.07 & 642.68 &  640.11 &  0.0577 & 0.7495\\
		Exponential &  & 0.1012 & 658.04 & 660.04 & 662.65 & 660.08 & 0.1630 &  0.0050\\
		\hline \hline

	\end{tabular}
\end{table}

\noindent 

\noindent Now the formal goodness of fit test is applied in order to verify which distribution performs better to the data. \textbf{}

\section{ Conclusion}

\noindent This paper studied the well--established and widely used Lindley distribution and its generalizations. The proposed distribution is discussed with various statistical properties and also with application on real data set. It performs better than various other distribution with different number of parameters. Generalization proposed here with its parameters can be used to handle various real data setswith complex structure. It can be also used to develop various new probability models to explain real phenomena.

 \begin{reference}

\item[] Abouammoh, A. M., Alshangiti, A. M., and Ragab, I. E. (2015). A new generalized Lindley distribution, {\em Journal of Statistical computation and simulation}, {\bf 85(18)}, 3662--3678.
\item[] Ali S. (2015). On the bayesian estimation of the weighted Lindley distribution. {\em Journal of Statistical computation and simulation}, {\bf 85(5)} : 855–880.
\item[] Alkarni, S. H. (2015). Extended power Lindley distribution: A new statistical model fornon-monotone survival data, {\em European journal of statistics and probability}, {\bf 3(3)}, 19--34.
\item[] Bakouch, H. S., Al-Zahrani, B. M., Al-Shomrani, A. A., Marchi, V. A., and Louzada, F. (2012). An extended Lindley distribution, {\em Journal of the Korean Statistical Society}, {\bf 41}, 75--85. 
\item[] Barlow, R.E. and R. Campo (1975). Total time on test processes and applications to failure data analysis. In: R.E. Barlow, J. Fussell and N.D. Singpurwalla, Eds., {\em Reliability and Fault Tree Analysis, SIAM, Philadelphia}, PA, 451-481.
\item[] Barreto-Souza W and Bakouch Hs. (2013). A new lifetime model with decreasing failure rate.Statistics: {\em A Journal of Theoretical and Applied Statistics}, {\bf 47(2)}: 465–476.
\item[]Bonferroni CE (1930) Elementi di Statistica Generale. Seeber. Firenze
\item[] Borah M and Begum Ra. (2002). Some properties of Poisson-Lindley and its derived distributions. {\em Journal of the Indian Statistical Association},{\bf 40(1)}: 13–25.
\item[] Borah M and Deka Na. (2001). A study on the inflated Poisson Lindley distribution. {\em Journal of the Indian Society of Agricultural Statistics}, {\bf 54(3)}: 317–323.
\item[] Borah M and Deka Na. (2001). Poisson-Lindley and some of its mixture distributions. {\em Pure and Applied Mathematika Sciences}, {\bf 53(1-2)}: 1–8.
\item[] Cakmakyapan, S., and Gamze, O. Z. E. L. (2016). The Lindley family of distributions: Properties and applications, {\em Hacettepe Journal of Mathematics and Statistics}, {\bf 46(6)}, 1113--1137. 
\item[] Elbatal, I., and Aryal, G. (2013). On the transmuted additive Weibull distribution, {\em Austrian Journal of Statistics}, {\bf 42(2)}, 117--132.
\item[] Ghitany Me, Al-Mutairi Dk and Nadarajah S. (2008). Zero-truncated Poisson-Lindley distribution and its application. {\em Mathematics and Computers in Simulation}, {\bf 79(3)}: 279–287.
\item[] Ghitany Me and Al-Mutari Dk. (2008). Size-biased Poisson-Lindley distribution and its application. {\em METRON – International Journal of Statistics}, {\bf 66(3)}: 299–311.
\item[] Ghitany Me, Alqallaf F, Al-Mutairi Dk and Husain Ha. (2011). A two-parameter weighted Lindley distribution and its applications to survival data. {\em Mathematics and Computers in Simulation},{\bf 81}: 1190–1201.
\item[] Ghitany, M. E., Al-Mutairi, D. K., Balakrishnan, N., and Al-Enezi, L. J. (2013). Power Lindley distribution and associated inference, {\em Computational Statistics and Data Analysis}, {\bf 64}, 20--33.
\item[] Ghitany, M. E., Atieh, B., and Nadarajah, S. (2008). Lindley distribution and its application, {\em Mathematics and computers in simulation}, {\bf 78(4)}, 493--506.
\item[] Gómez-Déniz, E., and Calderín-Ojeda, E. (2011). The discrete Lindley distribution: properties and applications, {\em Journal of Statistical Computation and Simulation},{\bf 81(11)}, 1405--1416.
\item[] Gupta, R. D., and Kundu, D. (2001). Generalized exponential distribution: different method of estimations, {\em Journal of Statistical Computation and Simulation}, {\bf 69(4)}, 315--337.
\item[] Gupta, R.D. and Kundu, D. (1999): Generalized Exponential Distribution, {\em Austalian and NewZealand Journal of Statistics}, {\bf 41(2)}, 173 --188.
\item[] Ibrahim, M., Yadav, A. S., Yousof, H. M., Goual, H., and Hamedani, G. G. (2019). A new extension of Lindley distribution: modified validation test, characterizations and different methods of estimation, {\em Communications for Statistical Applications and Methods}, {\bf 26(5)}, 473--495. 
\item[] Irshad, M. R. (2017). New extended generalized Lindley distribution: Properties and applications, {\em Statistica}, {\bf 77(1)}, 33--52.
\item[] Lindley, D. V. (1958). Fiducial distributions and Bayes' theorem, {\em Journal of the Royal Statistical Society, Series B (Methodological)} {\bf 20}: 102--107.
\item[] Lindley, D.V.  (1965). {\em Introduction to Probability and Statistics from a Bayesian Viewpoint, Part II: Inference}, Cambridge University Press, New York.
\item[] Mahmoudi, E., and Zakerzadeh, H. (2010). Generalized poisson–Lindley distribution, {\em Communications in Statistics-Theory and Methods}, {\bf 39(10)}, 1785--1798.
\item[] Mazucheli, J., and Achcar, J. A. (2011). The Lindley distribution applied to competing risks lifetime data, {\em Computer methods and programs in biomedicine}, {\bf 104(2)}, 188--192. 
\item[] Murthy, D.N.P., Xie, M., Jiang, R. (2004). Weibull Models, ,
USA, {\em John Wiley and Sons}.
\item[] Mazucheli, J., \& Achcar, J. A. (2011). The Lindley distribution applied to competing risks lifetime data. Computer methods and programs in biomedicine, {\bf 104(2)}, 188–192.
\item[] Nadarajah, S., and Haghighi, F. (2011). An extension of the exponential distribution, {\em Statistics}, {\bf 45(6)}, 543--558.
\item[] Nadarajah, S., Bakouch, H. S., and Tahmasbi, R. (2011). A generalized Lindley distribution, {\em Sankhya B}, {\bf 73(2)}, 331--359.
\item[] Nedjar, S., and Zeghdoudi, H. (2016). On gamma Lindley distribution: Properties and simulations, {\em Journal of Computational and Applied Mathematics}, {\bf 298}, 167--174.
\item[] Oluyede, B. O., and Yang, T. (2015). A new class of generalized Lindley distributions with applications, {\em Journal of Statistical Computation and Simulation}, {\bf 85(10)}, 2072--2100.
\item[] Pararai, M., Warahena-Liyanage, G., and Oluyede, B. O. (2015). A new class of generalized Power Lindley distribution with applications to lifetime data, {\em Theoretical mathematics and applications}, {\bf 5(1)}, 53.
\item[] Ramos, P. L., and Louzada, F. (2016). The generalized weighted Lindley distribution: Properties, estimation, and applications, {\em Cogent Mathematics and Statistics}, {\bf 3(1)}, 1256022.
\item[] Rényi, A. (1961). On measures of entropy and information. In Proceedings of the Fourth Berkeley Symposium on Mathematical Statistics and Probability, Volume 1: Contributions to the Theory of Statistics. The Regents of the University of California.
\item[] Rezaei, S., Sadr, B. B., Alizadeh, M., and Nadarajah, S. (2017). Topp–Leone generated family of distributions: Properties and applications, {\em Communications in Statistics-Theory and Methods}, {\bf 46(6)}, 2893--2909. 
\item[] Roozegar, R., and Nadarajah, S. (2017). On a generalized Lindley distribution,  {\em Statistica}, {\bf 77(2)}, 149--157.
\item[] Sah, B. K. (2015). A Two-Parameter Quasi-Lindley Mixture of Generalised Poisson Distribution,  {\em International Journal of Mathematics and Statistics Invention}, {\bf 3(7)}, 1-5. 
\item[] Satheesh Kumar, C., and Jose, R. (2019). On Double Lindley Distribution and Some of its Properties, {\em American Journal of Mathematical and Management Sciences}, {\bf 38(1)}, 23--43.
\item[] Sankaran, M. (1970). The discrete poisson-Lindley distribution, {\em Biometrics}, 145--149.
\item[] Shanker, R., and Amanuel, A. G. (2013). A new quasi Lindley distribution, {\em International Journal of Statistics and systems}, {\bf 8(2)}, 143--156.
\item[] Shanker, R., and Hagos, F. (2015). Zero-truncated Poisson-Sujatha distribution with applications, {\em Journal of Ethiopian Statistical Association}, {\bf 24}, 55--63.
\item[] Shanker, R., Hagos, F. and Selvaraj, S. (2015). On Modeling of Lifetimes Data using Exponential and Lindley Distribution, {\em Biometrics and Biostatistics International Journal}, {\bf 2(5)}, 140--147.
\item[] Shanker, R., and Mishra, A. (2013). A two-parameter Lindley distribution, {\em Statistics in Transition new series}, {\bf 1(14)}, 45--56.
\item[] Shanker, R., and Mishra, A. (2014). A two-parameter Poisson-Lindley distribution, {\em International journal of Statistics and Systems}, {\bf 9(1)}, 79--85.
\item[] Shanker, R., Shukla, K. K., Shanker, R., and Tekie, A. L. (2017). A three-parameter Lindley distribution, {\em American Journal of Mathematics and Statistics}, {\bf 7(1)}, 15--26.
\item[]Shannon, C. E. (1948). A mathematical theory of communication. {\em The Bell system technical journal}, {\bf 27(3)}, 379--423.
\item[] Shukla, K. K., and Shanker, R. (2019). Shukla Distribution and its Application, {\em Reliability: Theory and Applications}, {\bf 14(3)}, 46--55.
\item[] Singh Brijesh P., Niraj K. Singh and Shweta Dixit. 2015. Estimation of Parameters of Inflated Poisson-Lindley Distribution for Adult Out-migration. {\em Journal of Institute of Science and Technology, Tribhuvan University}, {\bf 20(2)}: 6-10.
\item[] Singh Brijesh P., Shweta Dixit and Tapan Kumar Roy. 2016. On Some Compounded Poisson Distributions with Applications to the Pattern of Number of Child Deaths. {\em Journal of Probability and Statistical Sciences}, {\bf 14(2)}: 203-209.
\item[] Singh, Brijesh P. and Das, Utpal Dhar (2020). Some Statistical Properties of a Weighted Distribution and Its Application to Waiting Time Data, {\em Asian Journal of Probability and Statistics}, {\bf 9(1)}, 1--12.
\item[] Tomy, L. (2018). A retrospective study on Lindley distribution, {\em Biometrics and Biostatistics International Journal}, {\bf 7}, 163--169.
\item[] Torabi, H., Falahati-Naeini, M., and Montazeri, N. H. (2015). An extended generalized Lindley distribution and its applications to lifetime data, {\em Journal of Statistical Research of Iran JSRI}, {\bf 11(2)}, 203--222.
\item[] Weibull, W. (1951). Wide applicability, {\em Journal of applied mechanics}, {\bf 103(730)}, 293--297.
\item[] Zakerzadeh, H., and Dolatia, A. (2009).  Generalized Lindley Distribution, {\em Journal of Mathematical extension}, 13--25.
\item[] Zamani H and Ismail N. 2010. Negative Binomial-Lindley distribution and its application. {\em Journal of Mathematics and Statistics}, {\bf 6(1)}: 4–9.

\end{reference}

\end{document}